\documentstyle[12pt]{article}
\newcommand{\be}{\begin{equation}}
\newcommand{\ee}{\end{equation}}
\newcommand{\bea}{\begin{eqnarray}}
\newcommand{\eea}{\end{eqnarray}}
\newcommand{\nn}{\nonumber \\}
\newcommand{\p}[1]{(\ref{#1})}
\newcommand{\lb}{\label}

\begin{document}
\begin{titlepage}
\begin{flushright}
hep-th/0605211\\
May 2006
\end{flushright}
\vskip 0.6truecm

\begin{center}
{\Large\bf Gauging ${\cal N}{=}4$ Supersymmetric Mechanics}
\vspace{1.5cm}

{\large\bf F. Delduc$\,{}^a$, E. Ivanov$\,{}^b$,}\\
\vspace{1cm}

{\it a)ENS-Lyon, Laboratoire de Physique, 46, all\'ee d'Italie,}\\
{\it 69364 LYON Cedex 07, France}\\
{\tt francois.delduc@ens-lyon.fr}
\vspace{0.3cm}

{\it b)Bogoliubov  Laboratory of Theoretical Physics, JINR,}\\
{\it 141980 Dubna, Moscow region, Russia} \\
{\tt eivanov@theor.jinr.ru}\\

\end{center}
\vspace{0.2cm}
\vskip 0.6truecm  \nopagebreak

\begin{abstract}
\noindent We argue that off-shell dualities between $d{=}1$ supermultiplets with
different sets of physical
bosonic components and the same number of fermionic ones are related to gauging
some symmetries in the actions of the supermultiplets with maximal sets of physical
bosons.
Our gauging procedure uses off-shell superfields and so is manifestly supersymmetric.
We focus on ${\cal N}{=}4$ supersymmetric
mechanics and show that various actions of the multiplet ${\bf (3,4,1)}$ amount
to some gauge choices
in the gauged superfield actions of the linear or nonlinear ${\bf (4,4,0)}$
multiplets.
In particular, the conformally invariant ${\bf (3,4,1)}$ superpotential
is generated by the Fayet-Iliopoulos term of the gauge superfield. We find
a new nonlinear variant of the multiplet ${\bf (4,4,0)}$, such that its
simplest superfield action produces the most general 4-dim hyper-K\"ahler
metric with one triholomorphic isometry as the
bosonic target metric. We also elaborate on some other instructive examples
of ${\cal N}{=}4$ superfield gaugings, including a non-abelian gauging
which relates the free linear ${\bf (4,4,0)}$ multiplet
to a self-interacting ${\bf (1,4,3)}$ multiplet.
\end{abstract}
\vspace{0.7cm}

\newpage

\end{titlepage}

\section{Introduction}

In recent years, the models of supersymmetric quantum mechanics, particularly those
with extended ${\cal N}{=}4$
and ${\cal N}{=}8, d{=}1$ supersymmetries, received considerable attention
(see e.g. \cite{rev} - \cite{GenSm} and
refs. therein). The interest in these theories is mainly
motivated by the hope that their thorough study could essentially
clarify the geometric and
quantum structure of some higher-dimensional ``parent'' theories
which yield the supersymmetric
mechanics models upon the appropriate reduction to one dimension.
Another source of interest
in such models lies in the fact that they can describe superextensions of some notable
quantum systems like integrable Calogero-Moser system, quantum Hall effect, etc.

The $d{=}1$ supersymmetric models are of interest also because of some specific features of
$d{=}1$ supersymmetry which have no direct analogs in the higher-dimensional cases.
One of such
peculiarities is that the $d{=}1$ analogs of some on-shell higher-dimensional
supermultiplets
are {\it off-shell} supermultiplets. For instance, the complex form of ${\cal N}{=}2, d{=}4$
hypermultiplet requires an infinite number of auxiliary fields for its off-shell
description, which
is naturally achieved within ${\cal N}{=}2$ harmonic superspace \cite{HSS,HSS1}.
On the other hand,
there exists an off-shell multiplet of ${\cal N}{=}4, d{=}1$ supersymmetry which
collects just 4
physical bosonic fields and 4 physical fermionic fields and no auxiliary fields at all
\cite{GR0}-\cite{IKL1}. This ${\bf (4,4,0)}$ multiplet cannot be obtained
by dimensional reduction from any off-shell $d{>}1$ multiplet. Its
${\cal N}{=}8$ analog is the off-shell ${\bf (8,8,0)}$ multiplet \cite{GR}
which also cannot be obtained by dimensional reduction from higher dimensions.

Another interesting feature of one-dimensional supersymmetries is the possibility to
obtain new off-shell multiplets by replacing the time derivative of some
physical bosonic field
by a new independent auxiliary field with the same transformation law. This phenomenon
was revealed in \cite{GR0,GR} and it was called there ``1D automorphic duality'' (see also \cite{PT}).
This duality manifests itself already in the simpler cases of ${\cal N}{=}1$
and ${\cal N}{=}2, d{=}1$ supersymmetries.
For instance, in the ${\cal N}{=}2$ case one can define
a real general superfield $\Phi(t, \theta, \bar\theta)$ with the off-shell
content ${\bf (1, 2, 1)}$ and a
complex chiral superfield $\varphi (t + i \theta\bar\theta, \theta)$ with the off-shell
content ${\bf (2, 2, 0)}\,$. The real or imaginary parts of the latter, say
\be
\Phi_+(t, \theta, \bar\theta) = \varphi(t + i\theta\bar\theta, \theta)
+ \bar\varphi(t - i\theta\bar\theta, \bar\theta), \label{I}
\ee
transforms just as $\Phi(t,\theta, \bar\theta)$, in which the auxiliary field appearing
as a coefficient before the highest-degree $\theta$-monomial, i.e.
$\theta \bar\theta A(t)\,$,
is substituted by
\be
A(t) = i\partial_t(\phi -\bar\phi)\,, \lb{III}
\ee
where $\phi = \varphi |_{\theta = \bar\theta = 0}\,$. Thus, if $\varphi$, $\bar\varphi$
enter some action
only through the combination \p{I}, one can treat the time derivative \p{III}
as an auxiliary field and
so dualize the action into an action of the ${\bf (1, 2, 1)}$ multiplet.
In refs. \cite{GR0,GR,PT} it was shown that this phenomenon is generic for $d{=}1$
supersymmetries. For instance,
in this way all off-shell ${\cal N}{=}4$ multiplets with 4 physical fermions, i.e.
${\bf (3, 4, 1)}$, ${\bf (2, 4, 2)}$, ${\bf (1, 4, 3)}$ and ${\bf (0, 4, 4)}$,
can be obtained from the ``master'' (or ``root'' \cite{Root,root}) multiplet ${\bf (4, 4,0)}$.
An ${\cal N}{=}8$ analog of the latter is the multiplet ${\bf (8,8,0)}$.
Alternatively, one could start from the
opposite end and substitute an auxiliary field by the time derivative of some new bosonic
field of physical dimension (as in \p{I}, \p{III}) thus enlarging the sets
of physical fields. The direct and inverse
procedures will be referred to as the ``reduction'' and ``oxidation'',
respectively\footnote{Though these terms, while applied to $d{=}1$ supersymmetry, are
also used in a different sense (see e.g. \cite{GenSm,T}), we hope that their usage
in the given context will not be misleading. Originally, the term ``oxidation''
was employed to denote the procedure inverse to the space-time dimensional reduction.
It seems natural to make use of the same nomenclature also as regards the target bosonic manifolds.}.

In \cite{GR0,GR1,GR,PT,Root} (see also \cite{GenSm}) these $d{=}1$ dualities were considered
at the linearized level. It was observed in
\cite{IL,IKL0,root} that there also exist nonlinear versions of these relations.
Namely, the multiplet
${\bf (3,4, 1)}$ can be constructed as a bilinear of two multiplets
${\bf (4,4,0)}$, so that the
auxiliary field of the former is not expressed as a simple time derivative.
Once again, this phenomenon
can be seen already in the ${\cal N}{=}2$ case: one can alternatively
construct a composite $\tilde{\Phi}$ as
\be
\tilde{\Phi}(t,\theta, \bar\theta) = \varphi (t + i\theta\bar\theta, \theta)
\bar{\varphi}(t - i\theta\bar\theta, \bar\theta) \quad \Rightarrow \quad \tilde{A} =
-i(\, \phi \partial_t\bar\phi - \bar\phi\partial_t\phi\,)\,. \label{II}
\ee

These dualities between different off-shell $d{=}1$ multiplets are very useful
since they allow one to
establish connections between the relevant invariant actions. For instance, the
superconformally invariant actions of the ${\cal N}{=}4$ multiplet ${\bf (4,4,0)}$
were found in \cite{IL}
by substituting, into the known superconformal actions of the multiplet
${\bf (3,4,1)}$ \cite{IKL0}, the
expression of the latter multiplet in terms of the former one.
It was also observed in \cite{Pol,IL}
that the general actions of the reduced multiplets are always
obtained from some subsets of the full variety of
the actions of multiplets one started with, and these subsets
are characterized by certain isometries
commuting with supersymmetry. In the above two ${\cal N}{=}2$ examples
the composite multiplets $\Phi_+$ and
$\tilde{\Phi}$ are invariant under the translational and rotational
abelian transformations of
the chiral superfields, viz.
\be
\mbox{(a)} \;\delta \varphi = i\lambda \qquad \mbox{and} \qquad \mbox{(b)} \;
\delta \varphi = i\lambda' \varphi\,, \lb{N2isom}
\ee
where $\lambda$ and $\lambda'$ are some real group parameters. Thus the actions
of the composite
superfields $\Phi_+$ and $\tilde{\Phi}$ form subclasses of the general $\varphi$ action,
such that they are invariant under isometries (\ref{N2isom}a) or (\ref{N2isom}b).

In the previous studies, the reduction procedure from multiplets with lesser
number of auxiliary
fields to those with the enlarged number and/or the inverse oxidation procedure were mostly
considered at the level of components, performing the relevant substitutions
of the auxiliary fields
``by hand''. In the case of nonlinear substitutions proceeding this way is
to some extent
an ``art''. The basic aim of the present paper is to put this procedure on a systematic
basis, staying at all steps within the manifestly supersymmetric framework of the
superfield approach. Using the formalism of harmonic ${\cal N}{=}4$ superspace
as most appropriate
for ${\cal N}{=}4, d{=}1$ supersymmetry \cite{IL}, we demonstrate that the path
leading from the
master ${\bf (4,4,0)}$ multiplet to its reduced counterparts in most of cases amounts to
gauging some isometries
of the off-shell ${\bf (4,4,0)}$ actions by a non-propagating ``topological''
gauge ${\cal N}{=}4$
superfield and choosing
some manifestly supersymmetric gauge in the resulting gauge-covariantized actions.
We consider both
the translational (shift) and rotational isometries. One of the notable outputs
of our analysis
is as follows. We construct a nonlinear version of the ${\bf (4,4,0)}$ multiplet
yielding in the
bosonic sector of the corresponding superfield action the general Gibbons-Hawking
(GH) ansatz
for hyper-K\"ahler 4-dimensional metrics with one triholomorphic isometry \cite{GH}
and accomplish the superfield gauging
of this isometry. As a result we obtain the general class of superfield actions
of the multiplet
${\bf (3,4,1)}$ which admit a formulation in the analytic harmonic
${\cal N}{=}4, d{=}1$ superspace \cite{IL}.
Another novel point is the interpretation of the conformally invariant
superpotential of the ${\bf (3,4,1)}$ multiplet as the Fayet-Iliopoulos term of the
gauge ${\cal N}{=}4$ superfield while gauging a rotational $U(1)$ isometry
of some linear
${\bf (4,4, 0)}$ multiplet. We also show that the actions of the so called
``nonlinear'' ${\bf (3,4,1)}$
multiplet \cite{IL,IKL1} can be reproduced by gauging an abelian target space
scale isometry in the
appropriate ${\bf (4,4,0)}$ actions. Our procedure is not limited solely
to the ${\bf (4,4,0)}$ actions.
We present the HSS formulation of the multiplet ${\bf (1,4,3)}$ and
demonstrate on the simple example
of the free action of this multiplet that the gauging of a shifting isometry of
the latter gives rise
to a superfield action of the fermionic multiplet ${\bf (0,4,4)}$. At last,
we give an example of
non-abelian gauging. We gauge the triholomorphic isometry $SU(2)_{PG}$ of the free
$q^{+ a}$ action (realized as rotations of the doublet index $a$) and,
as a result, recover
a non-trivial action of a self-interacting ${\bf (1,4,3)}$ multiplet.

In Sect. 2 we consider two simple ${\cal N}{=}1$ and ${\cal N}{=}2$ toy examples
of our gauging procedure.
We reproduce some actions of the multiplets ${\bf (0,1,1)}$ and ${\bf (1,2,1)}$
as special gauges
of the gauge-covariantized free actions of the multiplets ${\bf (1,1,0)}$ and ${\bf (2,2,0)}\,$.
The basic notions of ${\cal N}{=}4, d{=}1$ harmonic superspace are
collected in Sect. 3. In Sect. 4 we gauge shifting and rotational isometries
of the superfield actions of the multiplet ${\bf (4,4, 0)}$ defined by a linear harmonic
constraint. A generalization of these considerations to the case of the
nonlinear ${\bf (4,4,0)}$
multiplet producing the general GH ansatz in the bosonic sector
of the relevant action
is the subject of Sect. 5. Some further examples of the superfield
gauging procedure are presented
in Sect. 6.
\setcounter{equation}{0}
\section{${\cal N}{=}1$ and ${\cal N}{=}2$ examples}
The basic principles of our construction can be explained already
on the simple ${\cal N}{=}1, d{=}1$ example.
Let the coordinate set $(t, \theta)$ parametrize  ${\cal N}{=}1, d{=}1$ superspace and
$\Phi(t,\theta) = \phi(t) + \theta \chi(t)$ be a scalar ${\cal N}{=}1$ superfield
comprising the ${\cal N}{=}1$
supermultiplet ${\bf(1,1,0)}$. The invariant free action of $\Phi$ is
\footnote{Throughout the paper we do not care about
the dimension of the superfields involved, having in mind that the correct dimension
of the superfield actions can be ensured by
inserting appropriate normalization constants in front of them. Hereafter,
without loss of generality, we set such
constants equal to unity.}
\be
S_{N=1} = -i\int dt d\theta\, \partial_t\Phi D\Phi = \int dt \left[(\partial_t \phi)^2
+ i \chi \partial_t\chi\right], \lb{N1Act}
\ee
where
\be
D = \frac{\partial}{\partial \theta} + i \theta \frac{\partial}{\partial t}\,, \;\;
D^2 = i\partial_t\,,
\quad \int d\theta \theta = 1\,.
\ee
The action \p{N1Act} is invariant under constant shifts
\be
\Phi{}' = \Phi + \lambda\,. \label{N12}
\ee
Let us now gauge this shifting symmetry by replacing
$\lambda \;\rightarrow \; \Lambda(t,\theta)$
in \p{N12}.
To gauge-covariantize the action \p{N1Act}, we are led to introduce the
fermionic ``gauge superfield''
$\Psi(t,\theta) = \psi(t) + i\theta A(t)$ transforming as
\be
\Psi{\,}' = \Psi + D\Lambda\,,
\ee
and to substitute the ``flat'' derivatives in \p{N1Act} by the gauge-covariant ones
\be
S_{N=1}^{gauge} = -i\int dt d\theta \, \nabla_t\Phi {\cal D}\Phi\,,\label{N1cov}
\ee
with
\be
\nabla_t\Phi  = \partial_t \Phi + iD\Psi\,, \quad {\cal D}\Phi = D\Phi - \Psi\,.
\ee
Taking into account that $\Lambda(t,\theta)$ is an arbitrary superfunction,
one can choose the
``unitary gauge'' in \p{N1cov}
\be
\Phi = 0\,,
\ee
in which
\be
S_{N=1}^{\scriptsize{gauge}} = -\int dtd\theta\, D\Psi \Psi =
\int dt \left(i\psi \partial_t \psi + A^2 \right).\label{N1cov1}
\ee
This is just the free action of the ${\cal N}{=}1$ multiplet ${\bf (0,1,1)}$,
with $A(t)$ being
the auxiliary bosonic field. Thus we observe the phenomenon of transmutation
of the physical
bosonic field $\phi(t)$ of the ${\cal N}{=}1$ multiplet ${\bf (1,1,0)}$ into
an auxiliary bosonic field
$A(t)$ of another off-shell ${\cal N}{=}1$ multiplet, the ${\bf (0,1,1)}$ one.
This comes about
as a result of gauging a shift isometry of the action \p{N1Act} of the former multiplet.
In other words, the
${\bf (0,1,1)}$ action \p{N1cov1} is a particular gauge of the covariantized
${\bf (1,1,0)}$
action \p{N1cov}. This is analogous to the Higgs effect: the gauge
superfield $\Psi$ ``eats''
the Goldstone superfield $\Phi$ (``unitary gauge'') and as a result
becomes the standard (ungauged)
${\cal N}{=}1, d{=}1$ superfield comprising the irreducible ${\bf (0,1,1)}$
${\cal N}{=}1$ multiplet
$\psi(t), A(t)\,$.

One can come to the same final result by choosing a Wess-Zumino gauge, in which
$\Psi(t,\theta)$ takes the form
\be
\Psi_{WZ}(t,\theta) = i\theta\,A(t)\,, \quad \delta A(t) = \partial_t \lambda(t)\,,
\quad \lambda(t)
= \Lambda(t,\theta)|_{\theta =0}\,. \label{WZ0}
\ee
In this gauge \p{N1cov} becomes
\be
S_{N=1}^{\scriptsize{gauge}} = \int dt \left[(\partial_t \phi - A)^2
+ i \chi \partial_t\chi\right]
\ee
and the residual gauge freedom acts as an arbitrary shift of
$\phi(t), \;\phi{}'(t) = \phi(t)
+ \lambda(t)\,$. Fixing this freedom by the gauge condition $\phi (t) = 0$,
we once again come
to the ${\bf (0,1,1)}$ free action.

On this simplest example we observe that the phenomenon of interchangeability
between the physical
and auxiliary degrees of freedom in $d{=}1$ supermultiplets can be given
a clear interpretation in terms
of gauging appropriate isometries of the relevant superfield actions. After gauging,
some physical (Goldstone) bosons
become pure gauge and can be eliminated, while the relevant $d{=}1$ ``gauge fields''
acquire status of auxiliary fields.
This treatment can be extended to higher ${\cal N}$ $d{=}1$ supersymmetries. In the next
Sections we shall
discuss how it works in models of ${\cal N}{=}4$ mechanics.

But before turning to this, let us dwell on some unusual features of the
${\cal N}{=}1$ gauge multiplet above,
shared by its ${\cal N}{=}2$ and ${\cal N}{=}4$ counterparts.

Looking at the ``WZ'' gauge \p{WZ0} we see that only bosonic ``gauge''
field $A(t)$ remains in it,
without any fermionic partner, although we started from off-shell
${\cal N}{=}1$ supersymmetry.
However, no contradiction arises since, due to the residual gauge invariance,
one ends up
with $(0 + 0)$ off-shell degrees of freedom (locally). It is amusing that
${\cal N}{=}1$ supersymmetry
plus compensating gauge transformation needed to preserve the WZ gauge \p{WZ0}
yield $\delta_{SUSY} A(t) = 0$!
This is still compatible with the standard closure of $d{=}1$ supersymmetry
on $d{=}1$ translations. Indeed,
the latter act on $A(t)$ as $\delta_a A(t) = a\partial_tA(t) = \partial_t(aA)$,
i.e. this
group variation is a particular case of the residual gauge freedom. So one can say
that, modulo the
residual gauge transformations, the trivial supersymmetry transformations of $A(t)$
still have
one-dimensional translations as their closure, just similarly to what one observes
in WZ gauges
in higher dimensions. Though locally $\Psi(t,\theta)$ does not bring any new physical
degrees of freedom,
globally the field $A(t)$ surviving in the WZ gauge can differ from a pure gauge
$\sim \partial_t b(t)\,$,
and this property leads to surprising consequences when gauging isometries of
the $d{=}1$ supersymmetry actions.
This special feature suggests the name ``topological'' for  such $d{=}1$ gauge multiplets.

In the ${\cal N}{=}2$ case an analog of the gauge superfield $\Psi(t,\theta)$
is the real superfield
${\cal V}(t, \theta, \bar\theta)$
with the transformation law
\be
{\cal V}{\,}'(t,\theta, \bar\theta) =
{\cal V} (t,\theta, \bar\theta) + \frac{i}{2}\left[\Lambda(t_L, \theta) -
\bar\Lambda (t_R, \bar\theta) \right],
\quad t_L = t + i\theta\bar\theta\,, \;t_R = \overline{(t_L)} \label{N2gauge}
\ee
Though this transformation law mimics that of ${\cal N}{=}1, 4D$ gauge superfield,
in the WZ gauge only one bosonic field
survives, as in \p{WZ0}:
\be
{\cal V}_{WZ} = \theta\bar\theta A(t)\,. \label{WZ00}
\ee
So this gauge multiplet is also ``topological''. By making use of it,
one can study various gaugings of ${\cal N}{=}2$ supersymmetric mechanics models
and establish, in this way, the relations between
off-shell ${\cal N}{=}2$ multiplets ${\bf (2,2,0)}$, ${\bf (1,2,1)}$ and ${\bf (0, 2,2)}$.
As an example, let us consider the gauging of the $U(1)$ phase invariance
of the free action of the ${\bf (2,2,0)}$ multiplet.

As was said already, this multiplet is described by the chiral ${\cal N}{=}2$
superfield $\varphi(t_L, \theta) =
\phi(t_L) + \theta \psi(t_L)$, with the bilinear action
\be
S_{N=2}^{free} = -\int dt d^2\theta \left[ D\varphi(t_L,\theta)
\bar D\bar\varphi (t_R,\bar\theta) +
4c\, \varphi(t_L,\theta)\bar\varphi(t_R,\bar\theta) \right], \label{ChirAct1}
\ee
where
$$
D= \frac{\partial}{\partial \theta} + i\bar\theta\frac{\partial}{\partial t}\,,\;
\bar D= -\frac{\partial}{\partial \bar\theta} - i\theta\frac{\partial}{\partial t}\, ,
$$
where $c$ is a coupling constant, with dimension $t^{-1}$.
In \p{ChirAct1} the first term yields the standard free action of the involved fields,
$\sim \partial_t\phi\partial_t\bar\phi +\ldots $,
while the second piece is an ${\cal N}{=}2$ superextension of the WZ-type Lagrangian
$\sim i\left( \partial_t\phi\bar\phi -
\partial_t\bar\phi \phi\right)$ (the specific normalization of the WZ term
was chosen for further convenience).  One can also add some potential terms
as integrals over chiral and anti-chiral ${\cal N}{=}2$ superspaces, but we do not
consider them here.

The action \p{ChirAct1} possesses the evident $U(1)$ invariance
$\varphi{}' = e^{-i\lambda} \varphi\,, \;\bar\varphi{}' =
e^{i\lambda} \bar\varphi\,$ with a constant parameter $\lambda\,$. Under the
local version of these
transformations
\be
\varphi{\,}' = e^{-i\Lambda} \varphi\,, \quad \bar\varphi{\,}' =
e^{i\bar\Lambda} \bar\varphi\,,
\quad
\Lambda = \Lambda(t_L,\theta)\,,\;  \bar\Lambda = \bar\Lambda(t_R,\bar\theta) \label{LocLam}
\ee
the action ceases to be invariant and should be covariantized with the help of the
``topological'' gauge
superfield ${\cal V}$ with the transformation law \p{N2gauge}:
\be
S_{N=2}^{\scriptsize{gauge}} = -\int dt d^2\theta \left[ {\cal D}\varphi(t_L,\theta)
\bar{\cal D}\bar\varphi (t_R,\bar\theta)\, e^{2{\cal V}} +
4c\, \varphi(t_L,\theta)\bar\varphi(t_R,\bar\theta)\, e^{2{\cal V}}
+ 2\xi \,{\cal V} \right]. \label{ChirAct}
\ee
Here
\be
{\cal D} = D + 2 D\,{\cal V}\,, \quad \bar{\cal D} = \bar D + 2\bar D\, {\cal V}\,,
\ee
and we also added a Fayet-Iliopoulos (FI) term for ${\cal V}\,$, with coupling constant $\xi$ of the same dimension as the constant $c$. Using the
gauge freedom \p{LocLam},
one can choose the manifestly supersymmetric ``unitary'' gauge
\be
\varphi = 1\,.
\ee
Now the gauge freedom \p{LocLam} has been fully ``compensated''
and ${\cal V}$ becomes a general real ${\cal N}{=}2, d{=}1$ superfield with
the off-shell content ${\bf (1,2,1)}\,$.
The action \p{ChirAct} in this particular gauge becomes the specific action
of the latter multiplet:
\be
S_{N=2}^{W} = -\int dt d^2\theta \left( D W \bar DW + c\, W^2 + 2\xi\,
\ln W \right),\label{Wact0}
\ee
where we redefined $W = 2e^{\cal V}\,$. Thus we started
from the bilinear action of the multiplet ${\bf (2,2,0)}$,
gauged its rotational $U(1)$ symmetry and came to the action of the multiplet
${\bf (1,2,1)}$
with a non-trivial
superpotential as a result of the special gauge-fixing in the covariantized
${\bf (2,2,0)}$ action.

The superpotential in \p{Wact0} is generated by the WZ and FI terms in
the gauge-covariantized action,
and it is interesting to see what kind of scalar component potential
they produce. Expanding $W$ as
\be
W(t,\theta, \bar\theta) = \rho(t) + \theta \chi(t) - \bar\theta\bar\chi(t)
+ \theta\bar\theta \omega(t)\,,
\ee
and neglecting fermions, we find
\be
S_{N=2}^{W(\scriptsize{bos})} = \int dt \left[(\partial_t \rho)^2 +
\omega^2 - 2 c\,\rho\omega  - 2\xi\,\omega\rho^{-1} \right],
\ee
which, after eliminating the auxiliary field $\omega(t)$, is reduced
to the simple expression
\be
S_{N=2}^{W(\scriptsize{bos})} = \int dt \left[(\partial_t \rho)^2
- c^2\,\rho^2  - \frac{\xi^2}{\rho^{2}} -2c\xi\right],
\ee
Thus, gauging the free action of the ${\bf (2,2,0)}$ multiplet, we finally arrived
at the action of the multiplet
${\bf (1,2,1)}$ with a non-trivial scalar potential.
It describes a two-particle Calogero-Moser model modulo a trivial center of mass motion and a constant shift of the energy.

Note that the action \p{ChirAct1} possesses also a shift isometry
$\varphi{}' = \varphi + \omega\,$, $\omega $ being a complex
constant parameter. One can alternatively gauge this isometry,
and in the corresponding ``unitary'' gauge $\varphi = 0$
also recover a superfield action of ${\bf (1,2,1)}$ multiplet. It looks like \p{Wact0},
but with the term $\sim W$ instead
of $\ln W$ (this linear term can in fact be removed by a shift of $W$).

In the ${\cal N}{=}1$ and ${\cal N}{=}2$ examples considered, the inverse ``oxidation''
procedure is rather straightforward, though
non-unique in the ${\cal N}{=}2$ case. While in the  ${\cal N}{=}1$ action
\p{N1cov1} one should just choose $\Psi = D\Phi$, in the actions
of the ${\bf (1,2,1)}$ superfield $W$ one can substitute $W$ either as in \p{I},
or as in \p{II}, or even as
$W = \varphi^n \bar\varphi^n\,$, $n$ being some real number (for self-consistency,
one should assume that $\varphi$
has a constant background part). For instance, in order to reproduce the original
action \p{ChirAct1}, one should make in
\p{Wact0} the substitution $W \sim \varphi^{\frac12}  \bar\varphi^{\frac12}\,$.
Such a non-uniqueness of the oxidation procedure
as compared to the reduction one (which is just the gauging of some fixed isometry)
is a general phenomenon manifesting itself
also in models with higher ${\cal N}$.

So much for the toy examples and let us turn to the ${\cal N}{=}4$ case. In fact,
various ${\cal N}{=}1$ and ${\cal N}{=}2$
gauged models can be reproduced by passing to the ${\cal N}{=}1$ or ${\cal N}{=}2$
superfield formulations of the
${\cal N}{=}4$ models considered in the next Sections.

\setcounter{equation}{0}
\section{${\cal N}{=}4\,, \; d{=}1$ harmonic superspace}
\subsection{Definitions}
We start by recalling basics of ${\cal N}{=}4$, $d{=}1$ harmonic superspace (HSS) \cite{IL}.

The ordinary ${\cal N}{=}4, d{=}1$ superspace is defined as
\be
(t,\theta_{i},\, \bar\theta^{i})\,,\quad \bar\theta^{i} = \overline{(\theta_{i})}\,,\lb{N4}
\ee
where $t$ is the time coordinate and the Grassmann-odd coordinates
$\theta_{i},\, \bar\theta^{i}$ form doublets of the  automorphism
group $SU(2)_{A}$\footnote{The second automorphism $SU(2)$
of the ${\cal N}{=}4, d{=}1$ Poincar\'e supersymmetry is hidden in this notation;
it combines
$\theta_i$ and $\bar\theta_i\,$ into
a doublet.}. The ${\cal N}{=}4$ supertranslations act as
\be
\delta \theta_i = \epsilon_i\,, \quad \delta \bar\theta^i = \bar\epsilon^i\,,\quad
\delta t = i \left(\bar\theta^i\epsilon_i - \bar\epsilon^i\theta_i\right). \lb{N4susy}
\ee
The corresponding covariant derivatives are defined as
\begin{equation}
D^{i}=\frac{\partial}{\partial \theta_{i}}+i\bar\theta^{i}\partial_{t}\,,\;\;
\bar D_{i}=\frac{\partial}{\partial \bar\theta^{i}}+i\theta_{i}\partial_{t}\,, \quad
\{D^{i},\bar D_{j}\}=2i\delta^i_j\partial_{t}\,, \;\{D^{i}, D^{j}\}
= \{\bar D_{i},\bar D_{j}\} =0\,.
\end{equation}

${\cal N}{=}4, d{=}1$ HSS is defined as an extension of \p{N4} by the harmonics
$u^\pm_i \in SU(2)_A/U(1)\,$. The basic relations the harmonics satisfy
are $u^{-}_i = \overline{(u^{+i})}\,$,
$u^{+i}u^-_{i}=1\,$. The latter constraint implies the important completeness
relation
\be
u^+_iu^-_k - u^+_ku^-_i = \varepsilon_{ik}\,. \lb{CompL}
\ee
The coordinates of ${\cal N}{=}4, d{=}1$ HSS
in the {\it analytic} basis are
\begin{equation}
\left( t_{A}=t-i(\theta^+\bar\theta^-+\theta^-\bar\theta^+)\,,\;
\theta^\pm=\theta^{i}u^\pm_{i}, \bar\theta^\pm=\bar\theta^{i}u^\pm_{i}, \;u^\pm_k \right).
\end{equation}
The analytic subspace of HSS is defined as the coordinate subset
\begin{equation}
(t_{A},\theta^+,\bar\theta^+, u^\pm_{i})\equiv (\zeta, u).
\end{equation}
It is closed under the ${\cal N}{=}4$ supersymmetry \p{N4susy}.

In the ``central'' basis $(t, \theta_i, \bar\theta^k, u^{\pm i})$ we define
the harmonic derivatives
and the harmonic projections of spinor derivatives as
\be
D^{\pm\pm} = \partial^{\pm\pm} = u^{\pm}_i\frac{\partial}{\partial u^{\mp}_i}\,,
\quad D^\pm=u^\pm_{i}D^{i},\quad \bar D^\pm=u^\pm_{i}\bar D^{i}\,,. \lb{Cderiv}
\ee
Then in the analytic basis, the same spinor and harmonic derivatives read
\begin{eqnarray}
&& D^+=\frac{\partial}{\partial\theta^{-}},\,\, \bar D^+=
-\frac{\partial}{\partial\bar\theta^-},\,\,
D^{-}=-\frac{\partial}{\partial \theta^{+}}+2i\bar\theta^{-}\partial_{t_{A}},\,\,
\bar D^{-}=\frac{\partial}{\partial \bar\theta^{+}}+2i\theta^{-}\partial_{t_{A}}\,,\nn
&& D^{++}=\partial^{++}-2i\theta^+\bar\theta^+\partial_{t_{A}}
+\theta^+\frac{\partial}{\partial\theta^-}
+ \bar\theta^+\frac{\partial}{\partial\bar\theta^-}\,, \nn
&& D^{--}=\partial^{--}-2i\theta^-\bar\theta^-\partial_{t_{A}}
+\theta^-\frac{\partial}{\partial\theta^+}
+  \bar\theta^-\frac{\partial}{\partial\bar\theta^+}\,.
\end{eqnarray}
The precise form of the algebra satisfied by these derivatives
can be found using their explicit
expressions. The basic relations to be used in what follows are
\begin{eqnarray}
&&[D^{\pm\pm},D^\mp]=D^\pm,\,\, [D^{\pm\pm},\bar D^\mp]=\bar D^\pm,\,\,
\{D^+,\bar D^-\}=-\{D^-,\bar D^+\}=2i\partial_{t_{A}},  \nn
&& [D^{++},D^{--}]= D^{0}\,, \quad [D^0, D^{\pm\pm}] =
\pm 2 D^{\pm\pm}\,, \lb{DharmAl}\\
&& D^0 = u^+_{i}\frac{\partial}{\partial u^+_{i}}-u^-_{i}
\frac{\partial}{\partial u^-_{i}}+
\theta^+\frac{\partial}{\partial \theta^+}
+\bar\theta^+\frac{\partial}{\partial \bar\theta^+}
-\theta^-\frac{\partial}{\partial \theta^-}
-\bar\theta^-\frac{\partial}{\partial \bar\theta^-}\,.\label{Dalg}
\end{eqnarray}

The derivatives $D^+$, $\bar D^+$ are short in the analytic basis,
whence it follows that
there exist analytic ${\cal N}{=}4$ superfields $\Phi^{(q)}(\zeta, u)$
\be
D^+\Phi^{(q)} = \bar D^+\Phi^{(q)}=0 \quad \Rightarrow \quad \Phi^{(q)}
= \Phi^{(q)}(\zeta, u)\,,\lb{AnalPhi}
\ee
where $q$ is the external harmonic $U(1)$ charge. This Grassmann harmonic
analyticity is preserved
by the harmonic derivative $D^{++}$:
when acting on $\Phi^{(q)}(\zeta, u)$, this derivative yields an analytic
${\cal N}{=}4, d{=}1$ superfield of the
charge $(q+2)$.

Finally, the measures of integration over the full HSS and its analytic subspace
are given, respectively, by
\begin{eqnarray}
&& dudtd^4\theta=dudt_{A}(D^-\bar D^-)(D^+\bar D^+)=\mu_{A}^{(-2)}(D^+\bar D^+),\nn
&& \mu_{A}^{(-2)}=dud\zeta^{(-2)}=dudt_{A}d\theta^+d\bar\theta^+=dudt_{A}(D^-\bar D^-).
\end{eqnarray}

\subsection{${\cal N}{=}4\,, \; d{=}1$ multiplets in HSS}
Off-shell multiplets of ${\cal N}{=}4$, $d{=}1$ supersymmetry with four fermions
admit a concise description in
the HSS framework. The HSS formulations of the multiplets ${\bf (4,4,0)}$
and ${\bf (3,4,1)}$ are simpler, since these multiplets are described by analytic
superfields.

The multiplet ${\bf (4,4,0)}$ is described by a doublet analytic superfield
$q^{+a}(\zeta,u)$
of charge $1$
satisfying the non-dynamical harmonic constraint \footnote{For brevity,
in what follows we frequently
omit the index ``A'' of $t_A$.}
\begin{equation}
D^{++}q^{+a}=0 \quad \Rightarrow \quad q^{+a}(\zeta,u)=f^{ia}(t)u^+_{i}+\theta^+\chi^{a}(t)
+\bar\theta^+\bar\chi^{a}(t)+2i\theta^+\bar\theta^+\partial_t f^{ia}(t)u^-_{i}.\label{coq}
\end{equation}
The ${\bf (3,4,1)}$ multiplet is described by a charge $2$ analytic
superfield $W^{++}(\zeta,u)$
satisfying the harmonic constraint
\bea
D^{++}W^{++} = 0 \quad \Rightarrow \quad
W^{++}(\zeta,u) &=& w^{(ik)}(t)u^+_{i}u^+_{k}+\theta^+\psi^{i}(t) u^+_{i}
+\bar\theta^+\bar\psi^{i}(t) u^+_{i} \nn
&& + \,i\theta^+\bar\theta^+[F(t)+2\partial_t w^{(ik)}(t)u^+_{i}u^-_{k}].\label{Wconstr}
\eea
The Grassmann analyticity conditions together with the harmonic constraints \p{coq}
and \p{Wconstr} imply
that in the central basis
\bea
&& q^{+ a} = q^{ia}(t,\theta, \bar\theta)u^{+}_i\,, \;\;
D^{(i}q^{k)a} = \bar D^{(i}q^{k)a} = 0\,, \label{qconsC} \\
&& W^{++} = W^{(ik)}(t,\theta, \bar\theta)u^+_iu^+_k\,, \;\;
D^{(i}W^{kl)} = \bar D^{(i}W^{kl)} = 0\,. \lb{WconsC}
\eea

By analogy with the ${\cal N}{=}2, d{=}4$ HSS \cite{HSS}, one can also
introduce the ${\cal N}{=}4$, $d{=}1$ ``gauge multiplet''.
It is represented by a charge $2$ unconstrained analytic superfield
$V^{++}(\zeta,u)\,$ the gauge transformation of which reads in the abelian case
\begin{equation}
\delta V^{++}=D^{++}\Lambda\,,
\end{equation}
with $\Lambda(\zeta,u)$ being a charge zero unconstrained analytic
superfield parameter.
Using this gauge freedom, one can choose the Wess-Zumino gauge, in which
the gauge superfield becomes
\begin{equation}
V^{++}(\zeta,u)=2i(\theta^+\bar\theta^+)A(t),\quad\delta A(t)=
-\partial_{t}\Lambda_{0}(t),\,\,
\Lambda_0 = \Lambda(\zeta,u)\vert_{\theta = 0}\,. \lb{WZ}
\end{equation}
We observe here the same phenomenon as in the ${\cal N}{=}1$ and ${\cal N}{=}2$
cases of Sect. 2: the ``gauge''
${\cal N}{=}4, d{=}1$ multiplet locally carries $(0 + 0)$ degrees of freedom
and so it is ``topological''.
Globally the field $A(t)$ can differ from a pure gauge, and this feature
allows for its treatment
as an auxiliary field in the ``unitary'' gauges.

As in the ${\cal N}{=}2, d{=}4$ HSS \cite{HSS}, $V^{++}$ gauge-covariantizes
the analyticity-preserving
harmonic derivative $D^{++}\,$. Assume that the analytic superfield $\Phi^{(q)}$
is transformed under
some abelian gauge isometry as
\be
\delta_\Lambda \Phi^{(q)} = \Lambda\, {\cal I}\, \Phi^{(q)}\,,
\ee
where ${\cal I}$ is the corresponding generator. Then the harmonic derivative
$D^{++}$ is
covariantized as
\be
D^{++}\Phi^{(q)} \quad \Longrightarrow \quad {\cal D}^{++}\Phi^{(q)} =
(D^{++} - V^{++}\,{\cal I})\Phi^{(q)}\,.\lb{D++cov}
\ee
One can also define the second, non-analytic harmonic connection $V^{--}$
\be
{\cal D}^{--} = D^{--} - V^{--}\,{\cal I}\,, \quad \delta V^{--}
= D^{--} \Lambda\,. \lb{D--cov}
\ee
From the requirement of preserving the algebra of harmonic derivatives \p{DharmAl},
\be
[{\cal D}^{++}, {\cal D}^{--}] = D^0\,, \quad [D^0, {\cal D}^{\pm\pm}]
= \pm 2 {\cal D}^{\pm\pm}\,,
\ee
the well-known harmonic zero-curvature equation follows
\be
D^{++}V^{--} - D^{--}V^{++} = 0\,, \lb{Hzc}
\ee
which specifies $V^{--}$ in terms of $V^{++}$. One can also define
the covariant spinor derivatives
\be
{\cal D}^{-} = [{\cal D}^{--}, D^+] = D^- +(D^+V^{--})\, {\cal I}\,, \quad
\bar{\cal D}^{-} = [{\cal D}^{--}, \bar D^+] =
\bar D^- + (\bar D^+V^{--})\, {\cal I}\,, \lb{Spcov}
\ee
as well as the covariant time derivative
\be
\{D^+, \bar{\cal D}^- \} = 2i {\cal D}_t\,, \quad {\cal D}_t
= \partial_t  - \frac{i}{2}(D^+\bar D^+ V^{--})\,{\cal I}\,. \lb{Vectcov}
\ee
The important specific feature of the $d{=}1$ HSS is that the vector
gauge connection
\be
V \equiv D^+\bar D^+ V^{--},\, \quad \delta V
= -2i\partial_{t_A} \Lambda\,, \lb{Vconn}
\ee
is an analytic superfield, $D^+ V = \bar D^+ V = 0\,$,
so ${\cal D}_t$ preserves the analyticity.

In the WZ gauge \p{WZ} spinor gauge connections
are vanishing, while
\be
V \quad \Longrightarrow \quad 2i\,A(t)\,. \lb{WZV}
\ee

We will exploit these relations in Sect. 3.

\subsection{General $q^{+a}$ and $W^{++}$ actions}
We may write a general off-shell action for the ${\bf (4,4,0)}$ multiplet as
\begin{equation}
S_{q}=\int dudtd^4\theta \,{\cal L}(q^{+a}, q^{-b}, u^\pm) , \,\,
q^{-a}\equiv D^{--}q^{+a}. \lb{qact}
\end{equation}
After solving the constraint \p{coq} in the central basis of HSS,
the superfield $q^{\pm a}$
may be written in this basis as
$q^{\pm a}=u^{\pm}_{i}q^{ia}(t,\theta,\bar\theta)\,$, $D^{(i}q^{k) a}
= \bar D^{(i}q^{k)a} = 0$.
We then use the notation
\begin{equation}
L(q^{ia})=\int du\,{\cal L}(q^{+a}, q^{-b}, u^\pm),\quad S_{q}
=\int dtd^4\theta\, L(q^{ia}).
\end{equation}
The free action is given by
\begin{equation}
S_{q}^{\mbox{\scriptsize free}}
= -\frac{1}{4}\,\int dudtd^4\theta\,(q^{+a}q^-_{a})
=\frac{i}{2}\int dud\zeta^{(-2)}\,(q^{+a}\partial_t q^+_{a}). \lb{Freeq}
\end{equation}

The action \p{qact} produces a sigma-model type action in components,
with two time derivatives on
the bosonic fields and one derivative on the fermions.
One can also construct an invariant which
in components yields a Wess-Zumino type action,
with one time derivative on the bosonic fields
(plus Yukawa-type fermionic terms). It is given by
the following general integral over
the analytic subspace
\be
S^{WZ}_q = \int du d\zeta^{(-2)}\, {\cal L}^{+2}(q^{+a}, u^\pm)\,. \lb{WZq}
\ee

A general sigma-model type action for the ${\bf (3,4,1)}$
multiplet can be written as
\begin{eqnarray}
&& S_{W}=\int dudtd^4\theta\,{\cal L}(W^{++},W^{--},W^{+-},u^\pm)\,,
\label{Wact} \\
&& W^{--}= \frac{1}{2}(D^{--})^2W^{++}\,,\,\, W^{+-}=\frac{1}{2}D^{--}W^{++}.
\end{eqnarray}
The free action is given by
\begin{equation}
S_{W}^{\mbox{\scriptsize free}}=-\frac{1}{4}\int dudtd^4\theta\,
W^{++}(D^{--})^2W^{++}
=i\int dud\zeta^{(-2)}\,W^{++}\Pi_{2}^{--}W^{++},\label{truc}
\end{equation}
where the second-order differential operator
\begin{equation}
\Pi_{2}^{--}=D^{--}\partial_{t}+\frac{i}{2}D^-\bar D^-
= \partial^{--}\partial_{t_A} -
\frac{i}{2}\frac{\partial}{\partial \theta^+}
\frac{\partial}{\partial \bar\theta^+}\lb{Pi}
\end{equation}
preserves the analyticity:
\begin{equation}
[D^+,\Pi_{2}^{--}]=[\bar D^+,\Pi_{2}^{--}]=0\,,\;
[D^{++},\Pi_{2}^{--}]=(D_{0}-1)\partial_{t_{A}}
-\frac{i}{2}\left(\bar D^-D^+ -  D^-\bar D^+ \right)\,.\lb{PiComm}
\end{equation}

One can also define an analog of \p{WZq}:
\be
S^{WZ}_{W} = \int du d\zeta^{(-2)}\, {\cal L}^{+2}(W^{++}, u^\pm)\,.\lb{WZw}
\ee
In the component notation, it yields both the coupling to an external gauge
field background and the scalar
potential (plus appropriate fermionic terms).

\setcounter{equation}{0}
\section{Gauging isometries of the $q^+$ actions}
In the following, we shall consider two different types of isometries
admitting a realization on $q^{+a}$: shifts and rotations.

\subsection{Shift isometries}
Let us start with an example of shift isometry.
It will be convenient for us to equivalently
represent $q^{+a}$ by its harmonic projections:
\be
q^{+a} \quad \Longleftrightarrow \quad  L^{++} = q^{+ a}u^+_a\,, \;
\omega = q^{+ a}u^-_a\,.\lb{Lomega}
\ee
The constraints \p{coq} are rewritten in this notation as
\be
\mbox{(a)}\;D^{++}L^{++} = 0\,,
\quad \mbox{(b)}\;D^{++}\omega - L^{++} = 0\,. \lb{ConstrL}
\ee

Infinitesimal transformations of the isometry are
\begin{equation}
\delta q^{+a}=\lambda \,u^{+a}\quad \Longrightarrow
\quad \delta L^{++}=0\,,\quad
\delta\omega =\lambda\,, \lb{Shift}
\end{equation}
where $\lambda$ is a constant parameter. The free action of
the $q^{+a}$ superfields
\p{Freeq} is manifestly
invariant under this transformation,
because the integrand transforms to a total derivative
\begin{equation}
\delta S^{\mbox{\scriptsize free}}_{q}=
\frac{i}{2}\int dud\zeta^{(-2)}\,\partial_{t}(\lambda u^{+a}q^+_{a})=0.
\end{equation}

Let us now gauge this isometry
\begin{equation}
\lambda \;\Rightarrow \; \Lambda(\zeta,u)\,,\quad
\delta S^{\mbox{\scriptsize free}}_{q}
= i \int dud\zeta^{(-2)}\,\partial_t\Lambda\,(q^{+a} u^{+}_{a} )\,\neq 0\,.
\end{equation}
To gauge-covariantize the action, we apply the relations
\p{D++cov} - \p{Vectcov}. In the
case under consideration:
\be
{\cal I}\, q^{+ a} = u^{+a}\,.
\ee
Two steps are needed to make the theory gauge invariant.
\begin{enumerate}
\item{Covariantization of the constraint on $q^{+a}$ :
\begin{equation}
D^{++}q^{+a}=0 \quad \Rightarrow
\quad {\cal D}^{++}q^{+a}=D^{++}q^{+a}-V^{++}u^{+a} = 0\,.\lb{1constr}
\end{equation}
The equivalent form of this constraint in terms of
the superfields $L^{++}$ and $\omega$,
eq. \p{Lomega}, can be easily obtained by projecting on $u^\pm_i$:
\begin{equation}
\mbox{(a)}\;D^{++}L^{++} =0\,,\quad
\mbox{(b)}\;D^{++}\omega - V^{++} = L^{++}.\label{covc}
\end{equation}}
\item{Covariantization of the action $S^{\mbox{\scriptsize free}}_{q}\,.$\\
The gauge-invariant action is constructed using
the vector gauge connection \p{Vconn}
\begin{equation}
S^{\mbox{\scriptsize free}}_{q\,\mbox{\scriptsize (cov)}}=\frac{i}{2}
\int dud\zeta^{(-2)}
\left\{q^{+a}\partial_t q^+_{a}-iV(q^{+a}u^+_{a})+\xi V^{++}\right\}.\label{scc}
\end{equation}}
Here, we have taken into account that we are free to add
a FI term proportional to the arbitrary constant $\xi$. It is worth
noting that \p{scc} is not just
a replacement of $\partial_t$ in \p{Freeq} by the covariant derivative \p{Vectcov}.
This is related to the fact that
\p{Freeq} is invariant under the rigid isometry \p{Shift}
up to a total derivative in the integrand.
\end{enumerate}

The action (\ref{scc}) may be written as a general superspace integral
\begin{equation}
S^{\mbox{\scriptsize free}}_{q\,\mbox{\scriptsize (cov)}}=-\frac{1}{4}
\int dudtd^4\theta\left\{q^{+a}D^{--} q^+_{a}
-2V^{--}(q^{+a}u^+_{a})-2i\xi\theta^-\bar\theta^-
V^{++}\right\}. \label{cent}
\end{equation}
Using eq. (\ref{covc}a), it is convenient to replace the second term in (\ref{cent}) by
\begin{equation}
-2V^{--}(q^{+a}u^+_{a})\,\rightarrow -(D^{--})^2V^{++}(q^{+a}u^+_{a}).
\end{equation}
The equivalence of these two expressions may be shown by
substituting, in the second expression, $D^{--}V^{++}=D^{++}V^{--}$,
pulling $D^{++}$ to the left
using $[D^{++},D^{--}]=D^0$ and $D^0\,V^{--}=-2V^{--}$ and, finally,
integrating by parts with respect
to $D^{++}\,$, taking into account the constraint $D^{++}(q^{+a}u^+_a) = 0\,$.

Then, coming back to the analytic superspace notation, we get
\begin{equation}
S^{\mbox{\scriptsize free}}_{q\,\mbox{\scriptsize (cov)}}=\frac{i}{2}
\int dud\zeta^{(-2)}\left\{q^{+a}\partial_t q^+_{a}-2(q^{+a}u^+_{a})\Pi^{--}_2
V^{++}+ \xi\,V^{++}\right\},\label{chose}
\end{equation}
where the analyticity-preserving operator $\Pi^{--}_2$ was defined in \p{Pi}.

Instead of the WZ gauge, we can choose the unitary-type gauge
\begin{equation}
\omega=0,\label{ugauge}
\end{equation}
which may be reached since $\delta\omega=\Lambda$. Then, identifying
in this gauge $W^{++} \equiv L^{++} = q^{+ a}u^+_a$,
the basic constraints \p{covc} amount to
\begin{equation}
\mbox{(a)}\; D^{++}W^{++}=0\,,\quad
\mbox{(b)}\; V^{++}+W^{++}=0 \;\Rightarrow\;  V^{++}=-W^{++}.
\end{equation}
In the gauge (\ref{ugauge}), one has
\begin{equation}
q^{+a}\partial_t q^+_{a}=\partial_t\omega L^{++}-\omega\partial_t L^{++}=0,
\end{equation}
and the action (\ref{chose}) becomes
\begin{equation}
S^{\mbox{\scriptsize free}}_{q\,\mbox{\scriptsize (cov)}}={i}
\int dud\zeta^{(-2)}\left\{W^{++}\Pi^{--}_2W^{++}
-\frac{1}{2}\xi W^{++}\right\}.\label{uchose}
\end{equation}
Up to the FI term, we recover the free action (\ref{truc})
for the $W^{++}$ superfield.

Let us comment on the general sigma-model type action \p{qact}.
A simple analysis shows that
the gauge invariant subclass of such actions corresponds to the choice
\be
S_{q\,\mbox{\scriptsize (cov)}} = \int dudtd^4\theta \,
{\cal L}\left(L^{++}, D^{--}L^{++}, (D^{--})^2L^{++}, u\right). \lb{gaugeL}
\ee
In particular, the free gauge invariant Lagrangian in the central basis
(the first two terms in \p{cent})
can be written as (modulo purely analytic terms vanishing under
the full superspace integral)
\be
L^{++}(D^{--})^2L^{++}\,.
\ee
Any possible gauge invariant dimensionless quantity which can be constructed
from the gauge-variant projection
$\omega$ is reduced to one of the functional arguments in ${\cal L}$
in \p{gaugeL}, e.g.,
\be
D^{--}\omega - V^{--} = \frac{1}{2} (D^{--})^2L^{++}\,, \quad \mbox{etc.}
\ee
Thus in the unitary gauge \p{ugauge} we recover the general
${\bf (3, 4, 1)}$ action \p{Wact}.

Finally, the gauge-invariant subclass of the WZ terms \p{WZq} is
\be
S^{WZ}_{q\,\mbox{\scriptsize (cov)}} =
\int du d\zeta^{(-2)}\, {\cal L}^{+2}(q^{+a}u^+_a, u^\pm)\,. \lb{WZq1}
\ee
Note that the simplest choice
\be
{\cal L}^{+2}_0 = q^{+a}u^+_a = L^{++} \lb{L0}
\ee
gives a vanishing contribution in the ungauged case: indeed,
representing $u^+_a = D^{++}u^-_a$,
integrating by parts and taking account of the original constraint
\p{coq} yield zero. At
the same time, performing the same manipulations in the gauged case,
we obtain, due to
the modified constraint \p{1constr} (or \p{covc}), that
$$
{\cal L}^{+2}_0 \quad \Rightarrow \quad -V^{++} \neq 0\,
$$
i.e. \p{L0} is reduced to the FI term of $V^{++}\,$.

To summarize, passing to the gauge $\omega = 0$ in which $V^{++} = - L^{++}$
and identifying $W^{++} = L^{++}$
like in the case of the gauged free $q$-action, we found that all general
gauge-invariant $q$-actions \p{gaugeL},
\p{WZq1} became their $W$-counterparts defined in \p{Wact}, \p{WZw}.
Thus the general action of the ${\bf (3,4,1)}$
multiplet (the sum of the sigma-model and WZ pieces) is equivalent to the subclass of the
${\bf (4,4,0)}$ multiplet actions enjoying symmetry with respect to the gauged shift
isometry $q^{+ a} \Rightarrow q^{+a} + \Lambda u^{+a}\,$.
Of course, one could arrive at the same conclusion in the WZ gauge \p{WZ} for the harmonic
connection $V^{++}$. One can gauge away one physical bosonic field from
the multiplet ${\bf (4,4,0)}$ using the residual gauge freedom with
the parameter $\lambda_0(t)$,
after which the ``former'' gauge field $A(t)$ becomes the auxiliary field enlarging
the rest of ${\bf (3 + 4)}$ fields
to the off-shell ${\bf (3,4,1)}$ multiplet. The use of the unitary gauge $\omega = 0$ is
advantageous in that it preserves the standard realization of
${\cal N}{=}4$ supersymmetry at all steps.

\subsection{Rotational isometries}
We consider the following infinitesimal gauge transformations
\begin{equation}
\delta q^{+}_a=\Lambda{c_{a}}^{b}q^{+}_b,\quad c^{ab}=c^{ba},\,\, \overline{c^{ab}}
=c_{ab},\,\, c^{ab}c_{ab}\equiv c^2=2.\label{caram}
\end{equation}
Thus in this case
\be
{\cal I}q^{+}_a ={c_{a}}^{b}q^{+}_b\,.
\ee
The constraint (\ref{coq}) may be covariantized as
\begin{equation}
D^{++}q^{+a}+ V^{++}{c^{a}}_{b}q^{+b}=0.\label{ccovb}
\end{equation}
The simplest gauge invariant action is obtained by covariantizing
the free action \p{Freeq} and it reads
\begin{equation}
S_{q\mbox{\scriptsize (cov)}}^{\scriptsize{free}}
=\frac{i}{2}\int dud\zeta^{-2}\{q^{+a}\partial_t q^+_{a}
+\frac{i}{2}\,Vq^{+a}c_{ab}q^{+b}+\xi V^{++}\} \label{actg}
\end{equation}
(where we also added an independent FI term). It turns out convenient
to parametrize the superfields $q^{+a}$
as in eqs. (6.65), (6.66) of \cite{HSS1}
\begin{equation}
q^+_{a}=(c_{ad}+i\epsilon_{ad})(u^{+d}-ig^{++}u^{-d})e^{-i\frac{\omega}{2}}+
(c_{ad}-i\epsilon_{ad})(u^{+d}+ig^{++}u^{-d})e^{+i\frac{\omega}{2}}, \label{redef1}
\end{equation}
The constraint \p{ccovb} can be easily rewritten in these variables.
To simplify things, note that the
transformation (\ref{caram}) amounts to $\delta\omega= 2\Lambda$, so we can
again choose the supersymmetry-invariant gauge $\omega=0$. In this gauge
\begin{eqnarray}
&& q^{+a}=2({c^{a}}_{d}u^{+d}+g^{++}u^{-a})
= 2(c^{++}+g^{++})u^{-a}-2c^{+-}u^{+a}\,, \nn
&& (q^+u^+)=-2(c^{++}+g^{++})\,, \quad (q^+u^-)=-2c^{+-} \,.\label{Proj}
\label{bof}
\end{eqnarray}
Here $c^{\pm\pm} = c^{ab}u^\pm_au^\pm_b\,,$ $c^{+-}
= c^{ab}u^+_{(a}u^-_{b)}$ and
\be
c^{++}c^{--} - (c^{+-})^2 = \frac12 c^2 = 1\,,
\ee
where the completeness relation \p{CompL} is used. We may use
the expressions \p{Proj}
in eq. (\ref{ccovb}) and project it on $u^+_{a}$
and $u^-_{a}$ to obtain the constraints
\begin{eqnarray}&&
\quad D^{++}g^{++}+ V^{++}g^{++}c^{+-}=0,\label{cgp}\\&&
\quad V^{++}(1+c^{--}g^{++})- g^{++}=0\,\Rightarrow\,
V^{++}=\frac{g^{++}}{1+g^{++}c^{--}}.\label{cvp}
\end{eqnarray}
Substituting (\ref{cvp}) into (\ref{cgp}),  we obtain the nonlinear harmonic
constraint
\begin{equation}
D^{++}g^{++}=-\frac{(g^{++})^2c^{+-}}{1+c^{--}g^{++}}\,.\label{ccgp}
\end{equation}
After going to the parametrization
\begin{equation}
g^{++}=\frac{l^{++}}{1+\sqrt{1+c^{--}l^{++}}}\,\leftrightarrow\,
l^{++}=(2+c^{--}g^{++})g^{++},\label{bofp}
\end{equation}
the harmonic constraint (\ref{ccgp}) simplifies to
\begin{equation}
D^{++}l^{++}=0.\label{sim}
\end{equation}
The gauge superfield $V^{++}$ in (\ref{cvp}) may now be expressed as
\begin{equation}
V^{++}= \frac{l^{++}}{(1+\sqrt{1+c^{--}l^{++}})\sqrt{1+c^{--}l^{++}}}.
\end{equation}
{}From the last of equations (\ref{bof}) and from (\ref{bofp}), we also find
\begin{equation}
q^{+a}c_{ab}q^{+b}\equiv{\cal J}^{++}=4(c^{++}+l^{++}),
\end{equation}
and, as a consequence of (\ref{sim}), we have
\be
D^{++}{\cal J}^{++}=0\,. \lb{Con1}
\ee
It is worth pointing out that eqs. \p{Con1} and \p{sim} are the direct
corollaries of \p{ccovb} and
so hold irrespective of any particular gauge fixing.

In the gauge $\omega=0$ that we are considering, the action (\ref{actg})
also simplifies.
One can show that the first term in the integrand in \p{actg},
i.e. $q^{+a}\partial_t q^+_{a}$, vanishes
up to a full time derivative. The second term takes a form similar
to that obtained
in the case of a shift isometry
\begin{eqnarray}
&& \frac{i}{4}\int dud\zeta^{(-2)}\{iVq^{+a}c_{ab}q^{+b}\}=
\frac{i}{2}\int dud\zeta^{(-2)}\{V^{++}\Pi^{--}_2{\cal J}^{++}\} \nn
&& = \,2i\int dud\zeta^{(-2)}\frac{l^{++}}{(1+\sqrt{1+c^{--}l^{++}})
\sqrt{1+c^{--}l^{++}}}\,\Pi^{--}_2(c^{++}+l^{++})\,.
\label{bon}
\end{eqnarray}
We now have a non trivial sigma model action for the ${\bf (3,4,1)}$
multiplet described
by the superfield $l^{++}$. Recall that we started from a free
$q^+$ action with the rotational isometry,
we gauged this isometry and arrived in a unitary-type gauge $\omega = 0$
to a non-trivial $\sigma$-model action for the ${\bf (3,4,1)}$ multiplet.
The kinetic term (\ref{bon}) may be written in the full superspace as
\begin{eqnarray}
\frac{1}{4}\int dudtd^4\theta\frac{1}{(\sqrt{1+l^{++}c^{--}})^3}\,
(D^{--}l^{++} D^{--}l^{++})\,.
\lb{ConfKin}
\end{eqnarray}

Finally, the gauge-fixed form of the FI term in the action (\ref{actg}) is
\begin{equation}
\frac{i}{2}\,\xi\int dud\zeta^{(-2)}\frac{l^{++}}{(1+\sqrt{1+c^{--}l^{++}})
\sqrt{1+c^{--}l^{++}}}\,. \label{FIDi}
\end{equation}
This is just the conformally-invariant potential term involving a coupling to
a Dirac monopole background \cite{IKL0,IL}. One can see how simple
the derivation of this
term is in the approach based on gauging a rotational isometry as compared to
the original derivation in \cite{IL}. There, this Lagrangian was restored
step by step from the
requirement of superconformal invariance. It would be interesting to elaborate
on the superconformal properties of the sigma model action \p{ConfKin} and compare it
with the conformally-invariant actions of $q^{+a}$ deduced in \cite{IL} in
the framework of ordinary ${\cal N}{=}4, d{=}1$ superspace.

It is worthwile to note that we could from the very beginning add to the action
\p{actg}
the gauge invariant action
\be
\sim \int dud\zeta^{(-2)} {\cal L}^{+2}(q^{+ a}c_{ab}q^{+ b}, u^\pm)\,,
\ee
which in components yields a $U(1)$ invariant coupling of $f^{ia}(t)$ to
some background
gauge potential \cite{IL}. In the gauge $\omega = 0$ this term, together with the
FI term of $V^{++}$, produce the most general superpotential for
the ${\bf (3,4,1)}$ multiplet
described by $l^{++}$ (or ${\cal J}^{++} = 4 (l^{++} + c^{++})$).
One can also start from
the most general sigma model $q^+$ action invariant under the rotational
isometry
considered. The corresponding gauge invariant Lagrangian is a function
of three possible
invariants
\be
{\cal J}^{++} = q^{+ a}c_{ab}q^{+ b}\,, \;\; {\cal J}^{+-}
= q^{+ a}c_{ab}q^{- b}\,, \;\;
{\cal J}^{--} = q^{- a}c_{ab}q^{- b} \quad (q^{- a} \equiv {\cal D}^{--}q^{+ a})\,.
\ee
In the gauge $\omega = 0$ the full gauge-covariantized $q^{+a}$ action
(a sum of the sigma-model and
WZ-like terms) is again reduced to the most general action of the
${\bf (3,4,1)}$ multiplet, as
in the case of a shift isometry. Note that the field redefinition \p{redef1}
relating the rotational
and shifting realizations of the same isometry (on the superfields
$q^{+a}$ and $(l^{++}, \omega)\,$,
respectively) is highly nonlinear, so the gauge-invariant actions which look very
simple
in the shift case can take a rather complicated form in the rotational case and
vice versa.
For instance, the Dirac monopole superpotential \p{FIDi} is just the FI term
of $V^{++}$,
while in the shift case it is a sum of the relevant FI term and a properly chosen
function ${\cal L}^{+2}(q^{+ a}u^{+}_a, u^\pm)$.

It is worth pointing out that the final ${\bf (3,4,1)}$ actions do
not ``remember'' from
which gauged $q^{+ a}$ action they originated. Therefore the inverse procedure
of ``oxidation'' of
these actions to the $q^+$ actions is highly non-unique. For instance, one
can substitute $W^{++}$ in the
general $W$ actions either by $q^+\cdot u^+$, or
${\cal J}^{++} = q^{+ a}c_{ab}q^{+ b}$, with $q^{+ a}$
in both cases satisfying the linear constraint \p{coq}. In this way one
recovers general $q^+$ actions
possessing either shift or rotational isometries \cite{IL} (actually,
these subclasses are
related to each other via the redefinition \p{redef1}). Moreover, the same
$W^{++}$ actions can be
``oxidized'' to the actions of another type of ${\bf (4,4,0)}$ multiplet,
the nonlinear one.

\setcounter{equation}{0}
\section{Nonlinear (4,4,0) multiplet and its gauging}
The standard ${\cal N}{=}4$ ${\bf (4,4,0)}$ multiplet is defined
by the linear constraints \p{coq} or \p{qconsC}.
Its general sigma model action always yields a conformally flat metric
in the bosonic
sector \cite{IL}. On the other hand, as was noticed for the first time
in \cite{UnP}, one can deform the
off-shell constraint \p{coq} by some nonlinear terms, so as to gain
a hyper-K\"ahler
metric in the bosonic sector of the appropriate superfield $q^{+a}$ action
(non-trivial hyper-K\"ahler metrics
are not conformally flat). This new possibility
was demonstrated in \cite{UnP} for the simple example of the Taub-NUT metric
(see also \cite{bks} where
the Eguchi-Hanson example was constructed via the $d{=}1$ version of the
HSS quotient construction \cite{giot}).
Here we generalize this deformation of the $q^{+a}$ constraints in such
a way that,
in the bosonic sector of the superfield action,
a general Gibbons-Hawking (GH) ansatz \cite{GH} for 4-dimensional hyper-K\"ahler
metrics with one triholomorphic
isometry is reproduced (with Taub-NUT and Eguchi-Hanson as particular cases).
After
this we gauge this isometry
with the help of the non-propagating ``topological'' $V^{++}$ and
as output obtain the most general subclass of the ${\bf(3,4,1)}$ multiplet
actions
admitting
a closed formulation in the analytic harmonic ${\cal N}{=}4, d{=}1$ superspace.
The corresponding 3-dimensional bosonic metric is conformally flat and
satisfies the three-dimensional
Laplace equation. It is just a quotient of the general GH metric
by the action of a $U(1)$ isometry.

\subsection{Hyper-K\"ahler ${\cal N}{=}4$ mechanics from a nonlinear
deformation of the
multiplet (4,4,0)}
Our starting point will be the quadratic $q^{+a}$ action
\begin{equation}
S=\frac{i}{2}\int du d\zeta^{(-2)}(q^{+a}\partial_t q^+_{a})\,.\label{act11}
\end{equation}
As we saw above, the action \p{act11} gives rise to the free dynamics for
the ${\bf (4,4, 0)}$ multiplet, if the latter is defined by the
linear constraints \p{coq}.
But the same action produces a non-trivial sigma model if
the superfield $q^{+a}$ is required to
satisfy the nonlinear constraint
\begin{equation}
D^{++}q^{+a}+u^{+a}{\cal L}^{+2}(q^+\cdot u^+,u^\pm)=0\,,
\quad q^+\cdot u^+ \equiv q^{+a}u^+_a\,,  \label{cons1}
\end{equation}
where ${\cal L}^{+2}$ is an arbitrary charge $+2$ function of its arguments.
These constraints have been
chosen in such a way that they are invariant under the shift isometry \p{Shift}:
\begin{equation}
q^{+a}\,\,\longrightarrow\,\, q^{+a}+\lambda u^{+a}\,.\label{IsoSh1}
\end{equation}
In the case of generic ${\cal L}^{+2}$, no other symmetry
(besides ${\cal N}{=}4$ supersymmetry)
is respected by the theory. In particular, the ${\cal N}{=}4$
superalgebra automorphism
$SU(2)_A$ symmetry acting on the doublet indices of the harmonic
variables $u^{\pm i}$ is fully
broken due to the explicit harmonics in ${\cal L}^{+2}$.

Like in the examples considered in Sect. 3,  we may represent $q^{+a}$ by
its projections on $u^{+a}$ and $u^{-a}$
and rewrite \p{cons1} as the constraints for these projections
\be
(\mbox{a})\;\; D^{++}(q^{+}\cdot u^+) = 0\,, \quad (\mbox{b})\;\;
D^{++}(q^{+}\cdot u^-) - (q^{+}\cdot u^+)
+ {\cal L}^{+2}(q^+\cdot u^+, u^\pm) = 0\,. \label{consProj}
\ee
The action \p{act11} can be rewritten in terms of these projections
(modulo a total $t$-derivative) as
\be
S = i\int du d\zeta^{(-2)}(q^+\cdot u^+)\partial_t(q^+\cdot u^-)\,.
\label{Act22}
\ee

In what follows we will be interested in the bosonic field contents
of \p{consProj} and \p{Act22}.

The bosonic field content of the
superfield $q^{+a}$ is given by
\begin{equation}
q^{+a}=f^{+a}(t,u)+(\theta^+\bar\theta^+)A^{-a}(t,u)+\,\,\mbox{fermionic
terms}. \label{bosq}
\end{equation}
The nonlinear constraint \p{cons1}, like \p{coq}, leaves four bosonic and
four fermionic fields as the irreducible field content of $q^{+a}$ off shell
\footnote{The property that \p{cons1} are
{\it off-shell} constraints is the crucial difference with the case
of the ${\cal N}{=}2, d{=}4$ hypermultiplet in ${\cal N}{=}2, d{=}4$ HSS where
analogous relations are the equations of motion \cite{cmp,HSS1}.}.
One could explicitly
solve \p{cons1} and find the full expressions for all the components
in the expansion \p{bosq} in terms of
this irreducible ${\bf (4 + 4)}$ set.
Remarkably, in order to calculate the bosonic part of \p{Act22}
we do not need to know the full solution
of the component constraints.

After performing the Grassmann integration, the bosonic fields
defined in \p{bosq} appear
in the bosonic part of the action \p{act11} (or (\ref{Act22})) as
\begin{equation}
S_{bos}=\frac{i}{2}\int dtdu\,(f^{+a}(t,u)\partial_t A^{-}_a(t,u)
+A^{-a}(t,u)\partial_t
f^{+}_a (t,u)) = i \int dtdu\, A^{-a}(t,u)\partial_t
f^{+}_a (t,u)\,,\label{act2}
\end{equation}
where we integrated by parts with respect to $\partial_t$.
Using the completeness of the harmonic set $u^{\pm i}$, eq. \p{CompL},
we may write the fields
$f^{+a}$ and $A^{-a}$ as
\begin{equation}
f^{+a}(t,u)=-\varphi^{+2}(t,u)u^{-a}+v(t,u)u^{+a},\quad
A^{-a}(t,u)=\varphi^{-2}(t,u)u^{+a}+\varphi(t,u)u^{-a}.
\end{equation}
Being rewritten in terms of the harmonic projections
$\varphi^{\pm 2}$, $\varphi$ and $v$, the
action (\ref{act2}) becomes
\begin{equation}
S= i \int dtdu \,{\cal F}(t,u),\qquad {\cal
F}(t,u)=\varphi^{+2}\partial_t\varphi^{-2}+v\partial_t\varphi\,.\label{toc1}
\end{equation}
When all fermions are omitted, the bosonic part of the
constraint (\ref{cons1}) amounts to the following set of equations
for $\varphi^{\pm 2}$, $\varphi$ and $v$
\begin{eqnarray} &&
1)\,\,\partial^{++}\varphi^{+2}=0\,\Rightarrow\, \varphi^{+2}(t,u)
= x^{(ij)}(t)u^+_{i}u^+_{j} \equiv x^{++}\,, \\ &&
2)\,\,\partial^{++}v-x^{++}+{\cal L}^{+2}(x^{++},u)=0\,,
\label{cons4} \\ && 3)\,\,\partial^{++}\varphi+2i\partial_t
x^{++}=0\,\Rightarrow\,\varphi(t,u) =\varphi_{0}(t)-2i\partial_t
x^{+-}(t,u)\,,\,\partial^{++}\varphi_{0}=0\,, \label{cons3} \\ &&
4)\,\,\partial^{++}\varphi^{-2}+\varphi-2i\partial_t
v- \varphi \frac{\partial{\cal L}^{+2}}{\partial x^{++}}=0\,.
\label{cons2}
\end{eqnarray}
{}From now on, we use the notations
\begin{equation}
\frac{\partial{\cal L}^{+2}}{\partial x^{++}}\equiv{\cal L}\,,\quad
L\equiv\int du{\cal L},\quad A^{(ik)}=\int duu^{+(i}u^{-k)}{\cal
L}\,.
\end{equation}
We denote by $\varphi_{0}$ and $v_{0}$ those parts of $\varphi$ and
$v$ which are independent of harmonic variables. Then, taking
the harmonic integral
of both sides of (\ref{cons2}), it is easy to get
\begin{equation}
\varphi_{0}(1-L)-2i\partial_t v_{0}+2iA^{(ik)}\partial_t
x_{(ik)}=0\,\Rightarrow\,\varphi_{0}=\frac{2i}{1-L}[\partial_t
v_{0}-\partial_t x_{(ik)}A^{(ik)}]\,.\label{toca}
\end{equation}

Further, using the solution  (\ref{cons3}) for $\varphi$, we find
that
\begin{equation}
\int dtdu \,(v\partial_t\varphi)
=\int dtdu
[-\partial_t v_0\varphi_0-i(1-{\cal L})(\partial_t x^{++}\partial_t x^{--})]\,.
\label{act0}
\end{equation}
In this calculation we firstly used the solution for $\varphi(t,u)$,
then integrated
by parts with respect to $\partial^{++}$ and $\partial_t$ and finally
made use of eq. \p{cons4}.
In the same spirit, we can work out the second term
$\varphi^{+2}\partial_t\varphi^{-2}=x^{++}\partial_t\varphi^{-2}$ appearing
in the bosonic action (\ref{toc1})
\begin{eqnarray}
\int dtdu\, (x^{++}\partial_t\varphi^{-2})
&=& \int dtdu\,[-i(\partial_t x^{--}\partial_t x^{++})(1-{\cal L})+
\varphi_0\partial_t x^{(ik)}A_{(ik)} \nn
&&+ \,2i(\partial_t x^{+-}\partial_t x^{+-})(1-{\cal L})]\,.
\label{acta}
\end{eqnarray}
Here we have used the time derivative of eq. (\ref{cons4})
\begin{equation}
\partial^{++}\partial_t{v}=\frac{\partial}{\partial t}[x^{++} -{\cal
L}^{+2}(x^{++},u)]=\partial_t x^{++}(1-{\cal L})\,.\label{actb}
\end{equation}
Summing up (\ref{act0}) and (\ref{acta}), we obtain for the action
(\ref{toc1})
\begin{eqnarray}&&
S=i\int dtdu\,[-\partial_t v_{0}\varphi_{0}-2i(\partial_t x^{++}\partial_t x^{--}
-\partial_t x^{+-}\partial_t x^{+-})(1-{\cal L})
+\varphi_{0}\partial_t x^{(ik)}A_{(ik)}]\nonumber\\&&
=\frac{i}{2}\int dt\,[-\partial_t v_{0}\varphi_{0}-i\partial_t x^{(ik)}
\partial_t x_{(ik)}(1-{\cal L})
+\varphi_{0}\partial_t x^{(ik)}A_{(ik)}]\,. \label{Act}
\end{eqnarray}
To obtain the last expression, we made use of the relation
\be
\partial_t x^{++}\partial_t x^{--} - \partial_t x^{+-}\partial_t x^{+-}
= \frac12 \partial_t x^{ik}\partial_t x_{ik}\,,
\ee
which follows from the completeness condition \p{CompL}.
Substituting the expression in eq. (\ref{toca}) for $\varphi_{0}$ into \p{Act},
we find
\begin{equation}
S= \frac{1}{2}\int dt\left\{\frac{2}{1-L}[\partial_t v_{0}
-(\partial_t{x}^{(ik)}A_{(ik)})]^2+(1-L)\partial_t x^{(ik)}
\partial_t x_{(ik)}\right\}. \label{tocb}
\end{equation}

Introducing the notations
\begin{equation}
\vec A=-i(\vec\tau)_{(ik)}A^{(ik)}\,,\,\,
\vec x=\frac{i}{\sqrt 2}(\vec\tau)_{(ik)}x^{(ik)}\,,\,\,
\tilde v_{0}=\sqrt{2}v_{0}\,,
\end{equation}
where $\tau^a, \, a =1,2,3,$ are Pauli matrices, we cast (\ref{tocb})
into the form
\begin{equation}
S= \frac{1}{2}\int dt\left\{\frac{1}{1-L}(\partial_t{\tilde v}_{0}
+\partial_t{\vec x}\cdot\vec A)^2+(1-L)\partial_t{\vec x}\cdot
\partial_t{\vec x}\right\},\label{BosDist}
\end{equation}
which is just the $d{=}1$ pullback of the general
Gibbons-Hawking ansatz for four-dimensional hyper-K\"ahler metrics with
one triholomorphic $U(1)$ isometry.
The isometry is realized as a constant shift of $\tilde v_{0}$.
It is easy to check that, by definition,
\be
\Delta (1 -L) = \partial_{\tilde{v}_0}(1 -L) =  0\,,
\quad \vec \nabla \wedge \vec A = \vec \nabla (1 -L)\,,
\ee
which are general defining equations of the multicenter GH ansatz with the
potential $V \equiv 1 - L\,$.

For the case of Taub-NUT in the parametrization considered here ${\cal L}^{+2}$
is given by \cite{ARIVNI}
\begin{equation}
{\cal L}^{+2}=2\lambda\frac{(x^{++}-c^{++})}{(1+\sqrt{1+(x^{++}-c^{++})c^{--}})
\sqrt{1+(x^{++}-c^{++})c^{--}}}\,.
\label{LTN}
\end{equation}
Then
\begin{equation}
L=\int du\frac{\partial{\cal L}^{+2}}{\partial x^{++}}
=\lambda\int du\frac{1}{(\sqrt{1+(x^{++}-c^{++})c^{--}})^3}
=\lambda\frac{\sqrt{c^2}}{\sqrt{x^{ik}x_{ik}}}
=\sqrt{2}\lambda \frac{1}{\sqrt{x^{ik}x_{ik}}}\,,
\end{equation}
from which we obtain the typical Taub-Nut potential
\begin{equation}
1-L=1-\sqrt{2}\lambda\frac{1}{\vert\vec x\vert}\,. \label{TN1}
\end{equation}

It is interesting to compare this result with another form of the Taub-NUT metric,
corresponding to the realization
of the triholomorphic $U(1)$ isometry of the latter not as a shift
(like in the above parametrization) but
as a rotation. The corresponding ${\cal N}{=}4$ superfield action is still given by
\be
S_{TN} = \frac{i}{2}\int dud\zeta^{(-2)}\hat q^{+a}\partial_t{\hat q}_a^{+}\,,
\label{Act15}
\ee
while the constraint reads \footnote{This constraint is equivalent
to that of \cite{UnP} where
a particular form of the constant triplet $c^{(ab)}$ was used.}
\be
D^{++}\hat q^{+a} + f {\cal J}^{++} c^a_{\;\;b}\hat q^{+ b} = 0\,,
\quad {\cal J}^{++} \equiv \hat q^{+a}c_{ab}\hat q^{+b}\,.
\label{cons15}
\ee
where $f$ is a coupling constant and $c_{ab} = c_{ba}, c^2 = 2$.
The action and constraint
are invariant under the rigid $U(1)$ transformation embedded into the
$SU(2)_{PG}$ symmetry realized on the doublet indices $a, b$:
\be
\delta q^{+}_a = \lambda {c_a}^{\;\;b}q^{+}_b\,. \label{TranRot}
\ee
Note the ``conservation law'' which follows from \p{cons15}
\be
D^{++}{\cal J}^{++} = 0\,. \label{Cons}
\ee
The transformation law \p{TranRot} coincides with the rigid version of the
rotational isometry transformation \p{caram}
of the case of linear ${\bf (4,4,0)}$ multiplet, and \p{Cons} with \p{Con1}.

Now we wish to show that, using an appropriate change of variables, this particular
system can be
brought into the generic form of the ${\cal N}{=}4, d{=}1$ multicenter sigma model
as given above. This change
of variables was introduced in \cite{HSS} in the context of
${\cal N}{=}2, d{=}4$ hypermultiplets with
an infinite number of auxiliary fields and we already used it in Sect. 4.2.
It is given by eq. \p{redef1}.
Like in Sect. 4.2, we pass to the new analytic superfield $l^{++}$
which is related to $g^{++}$ in \p{redef1}
as
\be
g^{++} = \frac{l^{++}}{1 + \sqrt{1 + c^{--}l^{++}}}\,. \label{100b}
\ee
It is easy to check that
\be
{\cal J}^{++} = 4(c^{++} + l^{++})\,.
\ee
Then, as a corollary of  \p{Cons}, it follows that
\be
\quad D^{++}l^{++} = 0\,. \label{cons16}
\ee
The rotational isometry \p{TranRot}, in terms of the
superfields $\omega, l^{++}\,$, is realized as a shift
\be
\delta l^{++} = 0\,, \quad \delta \omega = 2\lambda\,.
\ee
A direct calculation shows that, modulo total derivatives,
\be
\hat q^{+a}\partial_t{\hat q}_a^{+} = -2 l^{++}\partial_t \omega\,. \label{2a}
\ee

Substituting \p{redef1} and \p{100b} into the constraint \p{cons15} while taking
into account its consequence
\p{cons16}, one finds that \p{cons15} also implies the following condition
on the superfield $\omega$:
\be
D^{++}\omega = 8 f (c^{++} + l^{++})
-2 \frac{l^{++}}{(1 + \sqrt{1 + c^{--}l^{++}})\sqrt{1 + c^{--}l^{++}}}\,.
\label{cons17}
\ee
Thus the original constraint \p{cons15}, being equivalently rewritten
in terms of the superfields
$\omega$ and $l^{++}$, amounts to two constraints \p{cons16} and \p{cons17}.

Now, combining $\omega$  and $L^{++} \equiv l^{++} + c^{++}$ into
a new $q^{+a}$ as
\be
q^{+ a} = u^{+ a} \frac12 \omega + u^{-a} L^{++}\,,\; (q^+\cdot u^+)
\equiv x^{++} = -L^{++}\,, \;
(q^+\cdot u^-) = \frac12 \omega\,,
\ee
we observe that the isometry \p{TranRot} is realized on this $q^{+a}$
just by shifts \p{IsoSh1},
so we can apply the whole reasoning which led from \p{act11} to \p{BosDist}.
Comparing \p{2a}, \p{cons17} with \p{Act22} and \p{consProj}, we find
\be
\hat q^{+a}\partial_t{\hat q}_a^{+} = 2q^{+a}\partial_t{q}_a^{+}
= -2 L^{++}\partial_t \omega +
\mbox{total $t$-derivative}  \label{qq}
\ee
and
\be
{\cal L}^{+2}_{TN} = (1+4f)x^{++}
- \frac{(x^{++} + c^{++})}{[1 + \sqrt{1 - c^{--}(x^{++} +c^{++})}]
\sqrt{1 - c^{--}(x^{++} + c^{++})}}\,.
\ee
The latter yields the potential
\be
V_{TN} = 1-L_{TN} = -4f + \frac{1}{\sqrt{2}\,\vert \vec x \vert}\,, \label{TN2}
\ee
which basically coincides with \p{TN1}. Taking into account the relation \p{qq},
we conclude that the action \p{Act15}
with the constraint \p{cons15} gives rise, up to some rescaling,
to the same Taub-NUT distance as \p{act11} with
the constraint \p{cons1} for the particular choice \p{LTN} of the
nonlinear function ${\cal L}^{+2} (q^+\cdot u^+, u)$
in \p{cons1}.

\subsection{Gauging}
We now wish to gauge the shift isometry of the action \p{act11}
and constraint \p{cons1}. The action becomes
\begin{eqnarray}&&
S=\frac{i}{2}\int dud\zeta^{(-2)}q^{+a}\partial_t{q}^+_{a}\,\rightarrow\,
S_{g}= \frac{i}{2} \int du d\zeta^{(-2)}\left\{q^{+a}
\partial_t{q}^+_{a}-iV(q^+\cdot u^{+})+ \xi\,V^{++}\right\}\nonumber\\&&
\Leftrightarrow S_{g}= \frac{i}{2}\int du d\zeta^{(-2)}\left\{q^{+a}\partial_t{q}^+_{a}-
2V^{++}\Pi^{--}_2(q^{+}u^{+})+ \xi\, V^{++}\right\}.\label{gact}
\end{eqnarray}
The last term in eq. (\ref{gact}), as in other cases, is a FI term. The
covariantized constraint reads
\begin{equation}
D^{++}q^{+a}-V^{++}u^{+a}+u^{+a}{\cal L}^{+2}(q^+\cdot u^+,u)=0.
\end{equation}
Projecting these equations on $u^+_{a}$ and $u^-_{a}$, we obtain
\begin{equation}
D^{++}(q^+\cdot u^+)=0\,,\,\,\, D^{++}(q^+\cdot u^{-})
-(q^+\cdot u^+)-V^{++}+{\cal L}^{+2}=0\,.
\end{equation}
In the gauge $q^+\cdot u^{-}=0$ these relations yield
\begin{equation}
V^{++}={\cal L}^{+2}(W^{++},u)-W^{++},\,\,\, W^{++} \equiv (q^+\cdot u^+),
\,\,\, D^{++}W^{++} = 0\,.
\end{equation}
The first term in the gauged action, $q^{+a}\partial_t{q}^+_{a}$, vanishes
in the gauge chosen.
The only superfield which remains is $W^{++}$ satisfying the constraint
$D^{++}W^{++}=0\,$.
The resulting action reads
\begin{equation}
S_{g}=\frac{i}{2}\int du d\zeta^{(-2)}\left\{ 2[W^{++}-{\cal L}^{+2}(W^{++},u)]
\Pi_2^{--}W^{++}+\xi\,({\cal L}^{+2}-W^{++})\right\}.\label{ActGene}
\end{equation}
The first piece in \p{ActGene} is the most general  sigma-model action for
the ${\bf (3,4,1)}$ multiplet which admits a representation in
${\cal N}{=}4$, $d{=}1$ harmonic superspace \cite{IL}. It is worth noting
that the whole nonlinearity
of this action originates from the function ${\cal L}^{+2}$ which initially
enters the nonlinear
$q^{+ a}$ constraint \p{cons1}. On the other hand, the property that
the same function specifies both
the sigma model and WZ parts is to some extent accidental. One could, from
the very beginning, add
to the action \p{act11} an independent WZ term invariant under \p{IsoSh1},
i.e. $\tilde{\cal L}^{+2}(q^+\cdot u^+, u)\,$.
This would have no impact on the structure of the sigma-model part, but would modify
the WZ term in the gauge-fixed action \p{ActGene} by adding
$\tilde{\cal L}^{+2}(W^{++}, u)$ and so would make
the WZ term fully independent of the sigma-model one.

It is worth saying a few words about the oxidation procedure in the present case.
One could forget about the origin of
the sigma-model part in the action \p{ActGene} and consider it just
as a convenient parametrization of a general analytic
superspace sigma-model action of the multiplet ${\bf (3,4,1)}$,
with ${\cal L}^{+2}(W^{++}, u)$ fully characterizing
such an action \cite{IL}. At the component level, such actions
are distinguished in that the target metric
satisfies the Laplace equation with respect to three target bosonic coordinates.
Like in the previous cases, one can oxidize
this $W^{++}$ action to the relevant subclass of actions of
the linear ${\bf (4,4,0)}$ multiplet, with one shift
or rotational isometry, and the conformally flat target metric
satisfying the four-dimensional analog of Laplace equation.
On the other hand, being aware of the existence of nonlinear
${\bf (4,4,0)}$ multiplet with the constraint \p{cons1}, one
can alternatively oxidize \p{ActGene} just to this new multiplet by
identifying $W^{++}$ with the relevant $(q^+\cdot u^+)$
and the nonlinear function in \p{cons1} with the function ${\cal L}^{+2}(W^{++}, u)$
from \p{ActGene}. After substituting
(\ref{consProj}b) into \p{ActGene}, integrating by parts with
respect to $D^{++}$
and using the last relation in \p{PiComm}, one arrives just at
the action \p{Act22}
which was the starting point in finding the HK $d{=}1$ sigma model with
the general GH
ansatz in Sect. 5.1. A component version of this superfield oxidation procedure
was described in a recent paper \cite{burg}.

Finally, we would like to note that
the transition to the ${\bf (3,4,1)}$ sigma model can perhaps be
even more clearly understood in the WZ gauge \p{WZ}
for $V^{++}\,$. In this gauge, the whole effect of covariantization
of the bosonic action \p{BosDist} amounts to the change
$\partial_t\tilde{v}_0 \;\Rightarrow \; \partial_t\tilde{v}_0 + \sqrt{2}\,A\,$.
The residual gauge freedom
acts as an arbitrary time-dependent shift of the field $\tilde{v}_0$, so we can further
fix it by the condition
$\tilde{v}_0 = 0\,$.  In this gauge, the first term in the covariantized bosonic
action \p{BosDist} basically becomes
just the square of the auxiliary field. In the absence of FI term
(i.e. for $\xi = 0$), the auxiliary field
fully decouples and the remaining bosonic distance is just
the second term in \p{BosDist}
$$
\sim (1-L)\partial_t{\vec x}\cdot \partial_t{\vec x}\,.
$$
It is conformally flat, with the conformal factor $(1-L)$ being
a general solution of the 3-dimensional Laplace equation.
If $\xi \neq 0$, the elimination of the auxiliary field induces
a non-trivial scalar potential and the coupling to an
abelian background gauge field
$$
-\frac{\xi^2}{4}(1 - L) + \frac{\xi}{\sqrt{2}}\,\partial_t{\vec x}\cdot {\vec A}\,.
$$

\setcounter{equation}{0}
\section{Some other  ${\cal N}{=}4$ gauged models}
Here we show that one physical bosonic field of the ${\bf (4,4,0)}$ multiplet
can be traded
for an auxiliary field in such a way that one ends up with
a nonlinear ${\bf (3,4,1)}$
multiplet introduced in \cite{IL,IKL1}. The corresponding isometry
to be gauged is some
non-compact scaling invariance. Also, we show how to describe the off-shell
multiplet
${\bf (1,4,3)}$ in ${\cal N}{=}4, d{=}1$ HSS and, by gauging the appropriate
shift symmetry of the
free action of this multiplet, to reproduce the action of the fermionic
multiplet ${\bf (0,4,4)}$
\cite{IL}. Finally, we give an example of non-abelian $SU(2)$ gauging
which directly
yields a ${\bf (1,4,3)}$ multiplet, starting from the ${\bf (4,4,0)}$ one.

\subsection{Nonlinear (3,4,1) multiplet from gauging (4,4,0)}
Besides the shift and rotational isometries, one can implement on
the superfield $q^{+a}$
also a rigid rescaling symmetry\footnote{It does not affect
the superspace coordinates
and so has no relation to the ordinary dilatations and superconformal group.
It can be interpreted
as dilatations in the target space.}
\be
\delta q^{+ a} = \lambda q^{+ a}\,.\lb{ScIso}
\ee
The linear constraint \p{coq} is obviously covariant under these rescalings.
One can
also define an invariant sigma-model type action with the Lagrangian which
can in general
depend on harmonics and all possible scale invariant ratios of the harmonic
projections
of $q^{+a}$ and $q^{-a}$, namely,
\bea
(q^{+}\cdot u^+)/(q^+\cdot u^-)\,, \; (q^-\cdot u^-)/(q^+\cdot u^-)\,, \;
(q^-\cdot u^+)/(q^+\cdot u^-)\,,
\eea
where, in order to avoid singularities, we are led to assume that the quantities
$q^{+}\cdot u^- = q^{+a}u^-_a$ and $q^-\cdot u^+ = q^{-a}u^+_a$ start with
constant parts
\footnote{These objects are
in a sense similar to the hypermultiplet conformal compensators in the
HSS formulation of the most general off-shell version of
${\cal N}{=}2, d{=}4$ Einstein supergravity \cite{HSS}.}(in principle, one
can also divide by non-singular linear combinations of $(q^+\cdot u^-)$
and $(q^-\cdot u^+)$).
The simplest such action is the scale-invariant analog of the free action \p{Freeq}.
It admits a concise
representation as an integral over the analytic HSS
\be
S^{\mbox{\scriptsize scale}}_q = -\frac14 \int dudtd^4\theta
\left(\frac{q^-\cdot u^+}{q^{+}\cdot u^-}\right) =
\frac{i}{2}\int du d\zeta^{(-2)}\left(
\frac{\partial_t q^{+ }\cdot u^+}{q^{+}\cdot u^-}\right).\label{Scact1}
\ee
It is straightforward to check that in the bosonic sector this action yields
the unique scale and
$SU(2)\times SU(2)$ invariant action \footnote{To calculate $G(f)$, one should
represent
$f^{ia} = -\varepsilon^{ia}f + f^{(ia)}$ whence $G(f) = \frac{1}{f^2}\int du
\frac{1}{[1 + (f^{(ia)}/f) u^{+}_{i}u^-_a]^2}\,$. The latter harmonic integral
can be directly computed,
e.g. by expanding the integrand in powers of $(f^{(ia)}/f) u^{+}_{i}u^-_a $
and doing the integrals
in each term.}
\be
S^{\mbox{\scriptsize scale}}_{q(\mbox{\scriptsize bos})}
= \frac12\int dt\, G(f ) \partial_t f^{ia}\partial_t f_{ia}\,, \quad
G(f) = \int du \frac{1}{(f^{ia}u^+_iu^-_a)^2}
= \frac{2}{(f^{ia}f_{ia})}\,.\label{BosAct}
\ee
The metric function $G(f)$ satisfies the four-dimensional harmonicity equation,
like in any
sigma-model $q^{+ a}$ action admitting a formulation in the analytic
$d{=}1$ HSS \cite{IL}.
One can also define the scale-invariant subclass of the general
analytic $q^+$ WZ terms \p{WZq}
\be
S^{WZ(\mbox{\scriptsize scale})}_q =
\int du d\zeta^{(-2)}
{\cal L}^{+2}\left(\frac{q^+\cdot u^+}{q^+\cdot u^-}, u^\pm\right). \lb{Scact2}
\ee

Let us now gauge the isometry \p{ScIso} as we did in Sect. 3, i.e.
by replacing the rigid parameter
$\lambda$ with an unconstrained analytic function $\Lambda(\zeta, u)$.
The gauge-covariant
generalization of \p{coq} reads
\be
(D^{++} - V^{++})q^{+a} = 0\,, \lb{GaugeSc}
\ee
or, in terms of the harmonic projections,
\be
\mbox{(a)} \;\; (D^{++} - V^{++})(q^+\cdot u^-) - (q^+\cdot u^+) = 0\,, \quad
\mbox{(b)} \;\; (D^{++} - V^{++})(q^+\cdot u^+) = 0\,. \lb{GaugeSc1}
\ee
The action \p{Scact1} can also be easily covariantized as
\bea
S^{\mbox{\scriptsize scale}}_{q(\mbox{\scriptsize cov})} &=& \frac14
\int dudtd^4\theta \left[\left(\frac{D^{--}q^+\cdot u^+}{q^{+}\cdot u^-}\right) -
V^{--} \left(\frac{q^+\cdot u^+}{q^{+}\cdot u^-}\right)\right]\nn
&=& \frac{i}{2}\int du d\zeta^{(-2)}\left[
\left(\frac{\partial_t q^{+}\cdot u^+}{q^{+}\cdot u^-}\right) -
\frac{i}{2} V \,\left(\frac{q^{+}\cdot u^+}{q^+\cdot u^-}\right)\right],
\label{Scact3}
\eea
where the analytic vector gauge connection $V(\zeta,u)$ was defined in \p{Vconn}.
The term \p{Scact2} is invariant
under both rigid and gauge transformations, so it does not require
any covariantization.

To reveal what the gauged model describes one can choose the WZ gauge \p{WZ}.
However,
like in the previous cases, we first use a gauge which preserves manifest
supersymmetry. The projection $\omega \equiv q^{+a}u^-_a$
is transformed as $\delta \omega = \Lambda \omega$.
Thus, taking into account that $\omega$ is assumed to start with
a constant, one can impose the gauge
$\omega =1\,$. Denoting $N^{++} = q^{+a}u^+_a$,
in this gauge the constraints \p{GaugeSc1} become
\be
\mbox{(a)} \;\; V^{++} = -N^{++}\,, \quad
\mbox{(b)} \;\; D^{++}N^{++} - V^{++}N^{++} = 0\; \; \Rightarrow \;
D^{++}N^{++} + N^{++}N^{++} = 0\,. \lb{GaugeSc2}
\ee
The only remaining superfield is the analytic superfield $N^{++}$,
and the constraint \p{GaugeSc2}
is just the defining constraint of the nonlinear ${\bf (3,4,1)}$
multiplet \cite{IL,IKL1}.
Thus the gauging of the scale isometry of the ${\bf (4,4,0)}$
multiplet leaves us with the nonlinear
${\bf (3,4,1)}$ multiplet, in contradistinction to the gauging
of the shift and rotational isometries
which leads to the {\it linear} ${\bf (3,4,1)}$ multiplet.

In the gauge $\omega = 1$ the first term in the second line of \p{Scact3}
becomes a total derivative, so \p{Scact3}
is reduced to
\be
S^{\mbox{\scriptsize scale}}_{q(\mbox{\scriptsize cov})} =
\frac{1}{4}\int du d\zeta^{(-2)}\, V\, N^{++}\,. \label{VNon}
\ee
It describes ${\cal N}{=}4$ superextension of some $d{=}1$ nonlinear sigma
model with 3-dimensional target manifold.
The Lagrangian in \p{VNon} is a function of $N^{++}$ since
$V = D^+\bar D^+ V^{--}$ and
$D^{++}V^{--} + D^{--} N^{++} = 0$ (recall eq. \p{Hzc}). This
action can be explicitly expressed in terms of $N^{++}$ in the central basis as
\be
S^{\mbox{\scriptsize scale}}_{q(\mbox{\scriptsize cov})} \sim
\int dt d^4\theta du dv\,
N^{++}(u)\,\frac{1}{(u^+\cdot v^+)^2}\, N^{++}(v)\,, \label{VNon1}
\ee
where both $N^{++}$ are given at the same superspace ``point''
$(t, \theta, \bar\theta)$ but at the different
harmonic sets $u^{\pm}_i$ and $v^{\pm}_i\,$. The harmonic Green
function $1/(u^+\cdot v^+)^2$ is defined in \cite{HSS1}.

To learn which $d{=}1$ sigma model underlies the action \p{VNon}, it is simpler to
make use of the WZ gauge in \p{GaugeSc1} and \p{Scact3}. For the bosonic action,
the effect of gauging
the scale invariance amounts to changing the time derivatives in \p{BosAct}
to the covariant one
\be
S \sim \int dt\, \frac{1}{(f^{ia}f_{ia})} \nabla_t f^{ia}\nabla_t f_{ia}\,,
\quad \nabla_t f^{ia}
=  (\partial_t + A)f^{ia}\,. \lb{bosScga}
\ee
Fixing the gauge with respect to the residual gauge freedom
$\delta f^{ia} = \lambda (t) f^{ia}$ in such a way that
\be
f^{ia}f_{ia} = 1\,, \label{ScG}
\ee
and eliminating the field $A(t)$ by its algebraic equation of motion
($A(t)$ fully decouples in the gauge
\p{ScG}), we can reduce \p{bosScga} to
\be
S \sim \int dt\, \partial_t f^{ia}\partial_t f_{ia}\,, \quad f^{ia}f_{ia} = 1\,.
\ee
It is just the action of $d{=}1$ sigma model for the principal chiral field,
i.e. for the coset
$S^3 \sim SO(4)/SO(3)\,$. Thus we started from the model the bosonic sector of
which is described by the scale-invariant sigma-model action \p{BosAct}
corresponding to a four-dimensional
target with a linearly realized $SO(4)$ symmetry. Then we gauged the target space
scale
invariance in this model and, as a result, gained the sigma model with
the homogeneous
three-dimensional target $\sim S^3$ with  half of $SO(4)\,$ nonlinearly realized.
One could na{\"\i}vely expect
that the gauging of the dilatations leads to the projective space
$RP^{3}=S^{3}/Z_{2}$
which is the space of
straight lines in $R^4$. However, the gauge fixing leaves untouched the ``large"
gauge transformations
$f^{ia}\rightarrow\,-f^{ia}$ and thus $S^3$ emerges\footnote{${\cal N}{=}4$
superextension
of the same bosonic
model was constructed in \cite{IKL1}, using the description of the nonlinear
${\bf (3,4,1)}$ multiplet
in the ordinary ${\cal N}{=}4, d{=}1$ superspace. Presumably, the superfield action
of \cite{IKL1} is related to
\p{VNon1} via a field redefinition.}. One can obtain
more general sigma models, with broken $SO(4)$, by gauging some other
scale-invariant $q^{+a}$
actions having no simple description in the analytic HSS. Also, it is
of clear interest to seek
a scale-invariant generalization of the nonlinear ${\bf (4,4,0)}$
constraint \p{cons1}
and to study the
relevant gauged models.

The superpotential term \p{Scact2} in the gauge $\omega = 1$ becomes
\be
S^{WZ}_{N} = \int du d\zeta^{(-2)} {\cal L}^{+2}\left(N^{++}, u^\pm\right).
\lb{Scact22}
\ee
The addition of the FI term $\sim \xi V^{++} = -\xi N^{++}$ in the present
case is not crucial,
since \p{Scact2} can include, prior to any gauging, the non-vanishing term
$\sim (q^{+a}u^+_a)/(q^{+a}u^-_a)\,$
which coincides with the FI term in the gauge $\omega = 1\,$.

\subsection{From (1,4,3) to (0,4,4)}
The off-shell multiplet ${\bf (1,4,3)}$ exists in two forms which differ
in the $SU(2)$
assignment of the auxiliary fields: they can be triplets with respect to
one or another automorphism $SU(2)$
groups forming the full $SO(4)$ automorphisms of ${\cal N}{=}4, d{=}1$
Poincar\'e supersymmetry \cite{IKLe,IKPa}.

One of these multiplets admits a simple description in the full
${\cal N}{=}4, d{=}1$ HSS.
It is represented
by the real superfield
$\Omega(x, \theta, \bar\theta, u)$ satisfying the constraints
\be
\mbox{(a)} \;\;D^+ \bar D^+ \Omega = 0\,, \quad \mbox{(b)}\;\;
D^{++}\Omega = 0\,.
\lb{Omega1}
\ee
In the analytic basis, eq. (\ref{Omega1}a) implies that $\Omega$
is linear
in the non-analytic coordinates,
\be
\Omega = \Sigma(\zeta, u) + i\left[\theta^- \Psi^+(\zeta, u)
+ \bar\theta^-\bar\Psi^+(\zeta, u)\right], \lb{Omega2}
\ee
while (\ref{Omega1}b) amounts to the following harmonic constraints
on the analytic superfunctions in \p{Omega2}
\be
\mbox{(a)}\;\;D^{++}\Psi^+ = D^{++}\bar\Psi^+ = 0\,, \quad \mbox{(b)}\;\;
D^{++}\Sigma +
i\left(\theta^+ \Psi^+ + \bar\theta^+\bar\Psi^+ \right) = 0\,. \lb{Omega3}
\ee
The general solution of \p{Omega3} is
\bea
&& \Psi^+ = \psi^iu^+_i + \theta^+ s + \bar\theta^+ r
+ 2i \theta^+\bar\theta ^+\partial_t \psi^i u^-_i\,, \nn
&& \bar\Psi^+ = -\bar\psi^i u^+_i - \theta^+ \bar r + \bar\theta^+ \bar s -
2i \theta^+\bar\theta^+ \partial_t \bar\psi^i u^-_i\,, \label{143Psi}\\
&& \Sigma = \sigma - i\theta^+ \psi^iu^-_i
+ i\bar\theta^+ \bar\psi^i u^-_i\,,\label{143}
\eea
where
\be
\mbox{Re}\,r = \partial_t \sigma\,. \lb{add}
\ee
The independent fields $\sigma(t), \psi^i(t), s(t), \mbox{Im}\, r(t)$
constitute an off-shell
${\bf (1,4,3)}$ multiplet.

The most general sigma-model type action of this multiplet is given by
the following
integral over the full
harmonic superspace
\be
S_{\Omega} =  \int du dt d^4\theta \, {\cal L}(\Omega, u^\pm)\,.
\ee
For our purposes we consider only its free part
\bea
S_{\Omega}^{free} &=& \int du dt d^4\theta\, \Omega^2
= \int du d\zeta^{(-2)} \,\Psi^+\bar\Psi^+ \lb{FreeG} \\
&=& \int dt \left[ (\partial_t\sigma)^2
- 2i \psi^i\partial_t \bar\psi_i + s\bar s + (\mbox{Im}\,r)^2\right].
\lb{FreeG1}
\eea
It is invariant under the constant shifts $\Omega \rightarrow
\Omega + \lambda$,
since the integral of $\Omega $
over the full superspace vanishes as a consequence of (\ref{Omega1}a).
Both constraints
\p{Omega1} are also manifestly invariant. Now let us gauge this isometry
by replacing, as usual,
$\lambda \rightarrow \Lambda(\zeta, u)$. The superfield action \p{FreeG}
(prior to passing to the
components) and the constraint (\ref{Omega1}a)
remain unchanged, while the harmonic constraint (\ref{Omega1}b) needs
the obvious covariantization
\be
\mbox{(\ref{Omega1}b)} \;\; \Rightarrow \;\; D^{++}\Omega - V^{++} = 0\,.
\label{OmegaG}
\ee
Clearly, one can choose the supersymmetric gauge
\be
\Sigma = 0\,,
\ee
in which the covariantized harmonic constraint \p{OmegaG} is reduced to
\be
\mbox{(a)}\;\;D^{++}\Psi^+ = D^{++}\bar\Psi^+ = 0\,, \quad \mbox{(b)}\;\;
V^{++} = i\left(\theta^+ \Psi^+ + \bar\theta^+\bar\Psi^+ \right). \lb{Omega4}
\ee
The solution of (\ref{Omega4}a) is just \p{143Psi} without the additional
condition \p{add}.
So we end up with the fermionic analog of the $q^+$ hypermultiplet,
the off-shell ${\bf (0,4,4)}$ multiplet formed by the fields $\psi^i(t), s(t), r(t)\,$.
The action \p{FreeG} in components becomes
\be
S_{\Psi}^{free} = \int du d\zeta^{(-2)} \,\Psi^+\bar\Psi^+
=\int dt \left( s\bar s + r\bar r - 2i \psi^i\partial_t \bar\psi_i \right).
\ee
One can also add the FI term $\sim \xi V^{++}$ which in the gauge considered
here yields
a ``potential'' component term
\be
S_{FI} = \frac{i}{2}\xi \int du d\zeta^{(-2)} V^{++} =
-\frac{\xi}{2} \int du d\zeta^{(-2)}
\left(\theta^+ \Psi^+ + \bar\theta^+\bar\Psi^+ \right) =
-\frac{\xi}{2}\int dt (r + \bar r)\,.
\ee

Less trivial examples can be obtained by considering a few independent
${\bf (1,4,3)}$ multiplets
and gauging some abelian isometries of the relevant sigma-model type superfield
actions. One of such
multiplets can always be traded for the ${\bf (0,4,4)}$ multiplet which will
non-trivially interact
with the remaining ${\bf (1,4,3)}$ multiplets. For instance, one can introduce
two superfields
$\Omega_1$ and $\Omega_2$, start with the scale invariant superfield
Lagrangian
${\cal L}(\Omega_1/\Omega_2, u)$ and gauge the scale invariance. Another
possibility is to
consider a complex $\Omega$ undergoing $U(1)$ transformations
$\delta \Omega = i\lambda \Omega$,
start with the $U(1)$ invariant Lagrangian ${\cal L}(\Omega\bar\Omega, u)$
and gauge this
$U(1)$ isometry. We plan to consider these and some other models elsewhere.

\subsection{An example of non-abelian gauging}
As our final example we consider non-abelian gauging of the ${\bf (4,4,0)}$
multiplet action. The covariantized action in a fixed gauge yields an action
of the ${\bf (1,4,3)}$
multiplet. Like in the previous example, we shall limit our consideration
to the gauging of the free action,
this time the free action of the $q^{+ a}$ multiplet \p{Freeq}, leaving
the consideration of more complicated
gauged systems for future study.

The action \p{Freeq} exhibits a manifest invariance under the global $SU(2)$
transformations which act
on the doublet index $a$ of $q^{+a}$ and so commute with ${\cal N}{=}4$
supersymmetry\footnote{By analogy
with the ${\cal N}{=}2, d{=}4$ case such $SU(2)$ symmetry can be called
``Pauli-G\"ursey'' $SU(2)$.}.
\be
\delta q^{+ a} = \lambda^a_{\;b}q^{+ b}\,, \quad \lambda^a_{\;a} = 0\,.
\ee

Let us gauge this symmetry by changing $\lambda^a_{\;b} \rightarrow
\Lambda^a_{\;b}(\zeta, u)$.
The constraint \p{coq} is covariantized to
\be
D^{++}q^{+ a} - {V^{++}}^a_{\;b}q^{+ b} = 0\,,\label{NAcoq}
\ee
where the traceless analytic gauge connection ${V^{++}}^a_{\;b}$ is transformed as
\be
\delta {V^{++}}^a_{\;b}=  D^{++}\Lambda^a_{\;b} + \Lambda^a_{\;c}{V^{++}}^c_{\;b}
- {V^{++}}^a_{\;c}\Lambda^c_{\;b}\,.
\ee
Using this freedom, one can pass to the WZ gauge as in the abelian case \p{WZ}
\bea
{V^{++}}^a_{\;b} = 2i \theta^+\bar\theta^+ A^{a}_{\,\;b}(t)\,, \;
\delta_r A^{a}_{\,\;b} =
-\partial_t{\Lambda_{(0)}}^a_{\;b}+ {\Lambda_{(0)}}^a_{\;c}A^{c}_{\;b}
- A^{a}_{\;c}{\Lambda_{(0)}}^c_{\;b}\,.  \label{WZ1}
\eea

The action \p{Freeq} is covariantized by changing
\be
\partial_t q^{+ a} \quad \Rightarrow \quad \nabla_t q^{+a}
= \partial_t q^{+a} -\frac{i}{2} V^a_{\,\;b}q^{+ b}\,,
\ee
where
\be
V^a_{\;\;b} = D^+\bar D^+ {V^{--}}^a_{\;b}\,, \quad
D^{++} {V^{--}}^a_{\;b} -  D^{--}{V^{++}}^a_{\;b}
- {V^{++}}^a_{\;c}{V^{--}}^c_{\;b} + {V^{--}}^a_{\;c}{V^{++}}^c_{\;b} =0\,.
\ee

In the present case we do not know a manifestly supersymmetric gauge, so we prefer
to deal with the WZ gauge \p{WZ1} in which
\be
{V^{--}}^a_{\;b} = 2i\theta^-\bar\theta^-\,A^{a}_{\;\;b}(t)\,,
\quad V^a_{\;\;b} = D^+\bar D^+ {V^{--}}^a_{\;b} =
2iA^{a}_{\;\;b}(t)\,.
\ee
In this gauge, the solution of the covariantized constraint \p{NAcoq}
is obtained from the solution \p{coq}
just by the replacement
\be
\partial_t f^{ia} \;\Rightarrow \;\nabla_t f^{ia} = \partial_t f^{ia}
+ A^a_{\;\;b}f^{i b}\,.
\ee
After substituting this covariantized solution for $q^{+a}$ into
the covariantization of the action \p{Freeq}
and performing there the Grassmann and harmonic integration,
we arrive at the following component action
\be
S^{bos}_{na} = \int dt\, \left( \nabla_t f^{ia}\nabla_t f_{ia}
- i \chi^a\nabla_t \bar\chi_a \right).\lb{Act8}
\ee
Splitting $f^{ia}$ as
\be
f^{ia} = \varepsilon^{ia} \frac{1}{\sqrt{2}}\,f + f^{(ia)}\,, \lb{Split}
\ee
and assuming that $f$ has a non-vanishing constant vacuum part,
$f = <f> + \ldots\,, <f> \neq 0$, one
observes that the symmetric part in \p{Split} can be fully gauged away
by the residual $SU(2)$ gauge
freedom
\be
f^{ia} \quad \Rightarrow \quad \varepsilon^{ia} \frac{1}{\sqrt{2}}\,f\,.
\ee
In this gauge, the action \p{Act8} becomes
\be
S^{bos}_{na} = \int dt\, \left[ (\partial_t f)^2  - i \chi^a\partial_t \bar\chi_a
+ f^2\,\frac{1}{2} A^{(ab)}A_{(ab)}
-i \chi_{(a}\bar\chi_{b)} A^{(ab)}  \right] \lb{Act9}
\ee
where the former gauge field $A^{(ab)}$ becomes a triplet of auxiliary fields.
So, gauging
the ``Pauli-G\"ursey'' $SU(2)$ symmetry of the {\it free} action of
${\bf (4,4,0)}$ multiplet
and choosing the appropriate gauge in the resulting covariantized action,
we arrived at the
action \p{Act9} which describes an {\it interacting} system of 1 physical
bosonic field $f(t)$,
the fermionic doublet $\chi^a(t)$ and the triplet of auxiliary fields $A^{(ab)}(t)$,
that is just the field content of off-shell
${\cal N}{=}4, d{=}1$ multiplet ${\bf (1,4,3)}\,$. The ${\cal N}{=}4$ supersymmetry of
the action \p{Act9} is guaranteed, since we
started from the manifestly supersymmetric action and just fixed some gauges in it.
The modified supersymmetry
transformations can be easily found, but we will not dwell on this. We only note that
after eliminating
the auxiliary field from \p{Act9}, the latter takes the following on-shell form
\be
S^{bos}_{na} = \int dt\, \left[ (\partial_t f)^2  - i \chi^a\partial_t \bar\chi_a  +
\frac{3}{8 f^2} (\chi^a\chi_a)(\bar\chi_a\bar\chi^a)\right]. \lb{Act10}
\ee
It remains to reveal whether we obtained in this way the standard linear ${\bf (1,4,3)}$
supermultiplet \cite{IKLe,IKPa}, or some new nonlinear version.

More general models of self-interacting ${\bf (1,4,3)}$ multiplet could be obtained via
the $SU(2)_{PG}$ gauging of more general sigma-model actions of $q^{+a}$.
It is impossible to construct
any non-trivial $SU(2)_{PG}$ invariant in the analytic HSS, besides
the free action \p{Freeq}.
In particular, any WZ term ${\cal L}^{+2}(q^+, u)$ necessarily breaks
$SU(2)_{PG}$. On
the other hand, in the full superspace one can easily construct
non-analytic $SU(2)_{PG}$
invariant
$q^{+ a}q^{-}_a$ and consider general $SU(2)_{PG}$ invariant sigma model
Lagrangians
as functions of this
invariant and explicit harmonics. We plan to consider the $SU(2)_{PG}$ gaugings of
such more general ${\cal N}{=}4, d{=}1$
actions elsewhere.

\setcounter{equation}{0}
\section{Conclusions}
In this article  we have shown, on a few instructive examples, that $d{=}1$
dualities between
supermultiplets with the same number of physical fermions but varying numbers
of physical and auxiliary
bosonic fields amount to gauging of the appropriate isometries of the relevant
superfield actions
by ``topological''
gauge multiplets. We mainly concentrated on the case of ${\cal N}{=}4$ supersymmetric
mechanics and considered
various reductions of the actions of the ${\bf (4,4,0)}$ multiplet to those
of the ${\bf (3,4,1)}$ multiplet.
Both linear and nonlinear versions of these multiplets were regarded.
As a by-product, we constructed a new
nonlinear ${\bf (4,4,0)}$ multiplet within a manifestly supersymmetric
superfield formulation. The target metric in
the bosonic sector of the simplest invariant action of this multiplet
is the general
4-dimensional HK metric with one triholomorphic isometry.

Although our consideration here was limited to ${\cal N}\leq 4$, the essence of
the method is expected to be the same for any ${\cal N}$,
and it goes in two steps. In a first step, one uses the gauge freedom
to get rid of
the whole of the topological gauge supermultiplet but the gauge field.
In this Wess-Zumino gauge, the gauge field does not
transform at all under supersymmetry. Further, what remains of the
gauge freedom is used to get rid of one physical bosonic field
in a unitary-type gauge. With this additional gauge-fixing,
the gauge field becomes an auxiliary field of the new (reduced) multiplet.
In practice, every time when this is possible it proves more advantageous
to use a manifestly supersymmetric gauge (if it exists) at each step of
the procedure.
The outcome naturally remains the same, but a clear merit of this alternative
gauge choice
is the opportunity to stay within the superfield approach.
While the direct (``reduction'') procedure is always fully specified
by the choice of isometry to be gauged and the form of the gauged action,
the inverse one (``oxidation'') is not unique:
the same reduced superfield action can be promoted to different particular
actions of the ``oxidized''
multiplet, depending on the choice of the substitution which expresses
the reduced  multiplet in terms of the ``oxidized'' one.

One of the possible further areas where our new approach could be efficiently
employed is ${\cal N}{=}8$
supersymmetric mechanics. There exists a plethora of various ${\cal N}{=}8$
multiplets \cite{BIKL}, as well as of associated mechanics
models (see e.g. \cite{GenSm} and refs. therein), and our procedure could probably
be very helpful in establishing
interrelations between these multiplets and models. There still remain some open
problems in the ${\cal N}{=}4$ case. For instance, it
would be of obvious interest to understand how to reproduce in our approach
the reductions of the ${\bf (4,4,0)}$ multiplet
to the off-shell multiplets ${\bf (2, 4, 2)}$. These reductions, both for
the linear \cite{FRa,IKLe} and nonlinear
\cite{IKL1} versions of the multiplet ${\bf (2, 4, 2)}$, were described
in \cite{root} at the component level. As one of
the conceivable ways of translating them into the superfield language,
it is natural to simultaneously gauge two mutually commuting isometries,
e.g. the rotational and scale isometries \p{TranRot} and \p{ScIso},
realized on the same superfield $q^{+ a}$.  Also, only the simplest
case of a nonabelian gauging which relates the multiplets ${\bf (4,4,0)}$
and ${\bf (1,4,3)}$ has been studied in this article. There are other
possibilities which still require to be investigated.

Finally, we would like to point out that a necessary condition of applicability
of our superfield gauging procedure is the commutativity of the isometries
to be gauged with the global supersymmetry (triholomorphicity). For isometries
which
do not meet this criterion (e.g. those belonging to R-symmetry groups)
a gauging seems
to be possible only within an extended framework of
non-dynamical superfield $d{=}1$ supergravities \cite{PD,BIS}.

\section*{Acknowledgments}
The work of E.I. was supported in part by the NATO grant PST.GLG.980302,
the RFBR grant 06-02-16684 and a grant of Heisenberg-Landau program. He thanks
Laboratoire de Physique, ENS-Lyon, and Universitat de Valencia - Departament
de Fisica Teorica, for the kind hospitality extended to him at the initial
and final stages of this study. We thank S.J.~Gates, Jr. for interest in the work
and valuable correspondence.

\end{document}